\documentclass[9pt,twocolumn,twoside]{osajnl}
\journal{ao}
\setboolean{shortarticle}{false}

\usepackage{caption}
\usepackage{subcaption}
\usepackage{booktabs}
\usepackage{stfloats}
\usepackage{color}
\usepackage{tikz}
\newcommand{\padinetal}{P18}
\title{Freeform three-mirror anastigmatic large-aperture telescope and receiver optics for CMB-S4}

\author[a]{Patricio A. Gallardo}
\author[b]{Roberto Puddu}
\author[c,d]{Kathleen Harrington}
\author[a,d,e]{Bradford Benson}
\author[a,c,d,f,g]{John Carlstrom}
\author[h]{Simon R. Dicker}
\author[i]{Nick Emerson}
\author[j,k]{Jon E. Gudmundsson}
\author[h]{Michele Limon}
\author[a,d,e,f,g]{Jeff McMahon}
\author[l]{Johanna M. Nagy}
\author[a,d]{Tyler Natoli}
\author[m,n]{Michael D. Niemack}
\author[o]{Stephen Padin}
\author[l]{John Ruhl}
\author[e]{Sara M. Simon}
\author[p]{the CMB-S4 collaboration  }
\affil[a]{Kavli Institute for Cosmological Physics, University of Chicago, Chicago, IL, USA}
\affil[b]{Instituto de Astrof\'isica and Centro de Astro-Ingenier\'ia, Facultad de F\'isica, Pontificia Universidad Cat\'olica de Chile, Santiago, Chile}
\affil[c]{High Energy Physics Division, Argonne National Laboratory, Argonne, IL, USA}
\affil[d]{Department of Astronomy and Astrophysics, University of Chicago, Chicago, IL, USA}
\affil[e]{Fermi National Accelerator Laboratory, Batavia, IL, USA}
\affil[f]{Department of Physics, University of Chicago, Chicago, IL, USA}
\affil[g]{Enrico Fermi Institute, University of Chicago, Chicago, IL, USA}
\affil[h]{Department of Physics and Astronomy, University of Pennsylvania, Philadelphia, PA, USA}
\affil[i]{Steward Observatory, The University of Arizona, Tucson, Arizona, USA}
\affil[j]{The Oskar Klein Centre, Department of Physics, Stockholm University, Stockholm, Sweden}
\affil[k]{Science Institute, University of Iceland, 107 Reykjavik, Iceland}
\affil[l]{Physics Department, Case Western Reserve University, Cleveland, OH, USA}
\affil[m]{Department of Physics, Cornell University, Ithaca, NY, USA}
\affil[n]{Department of Astronomy, Cornell University, Ithaca, NY, USA}
\affil[o]{California Institute of Technology, Pasadena, CA, USA}
\affil[p]{CMB-S4 collaboration: https://cmb-s4.org/team/}

\affil[*]{Corresponding author: pgallardo@uchicago.edu}

\begin{abstract}
CMB-S4, the next-generation ground-based cosmic microwave background (CMB) observatory, will provide detailed maps of the CMB at millimeter wavelengths to dramatically advance our understanding of the origin and evolution of the universe. CMB-S4 will deploy large and small aperture telescopes with hundreds of thousands of detectors to observe the CMB at arcminute and degree resolutions at millimeter wavelengths.
Inflationary science benefits from a deep delensing survey at arcminute resolutions capable of observing a large field of view at millimeter wavelengths. This kind of survey acts as a complement to a degree angular resolution survey. The delensing survey requires a nearly uniform distribution of cameras per frequency band across the focal plane. We present a large-throughput, large-aperture (5-meter diameter) freeform three-mirror anastigmatic telescope and an array of 85 cameras for CMB observations at arcminute resolutions, which meets the needs of the delensing survey of CMB-S4. A detailed prescription of this three-mirror telescope and cameras is provided, with a series of numerical calculations that indicate expected optical performance and mechanical tolerance.
\end{abstract}

\begin{document}
\maketitle
\section{Introduction}
\label{sec:introduction}
Advances in sensitivity of cosmic microwave background (CMB) observatories in the last decades and the rich science that benefits from these observations  have motivated the need for a large-scale CMB survey capable of delivering low noise maps of the cosmic microwave background at arcminute and degree angular scales, such as CMB-S4. The CMB-S4 science goals are broad; they include the search for primordial gravitational waves (a signature of early inflation), constraining dark energy, determining the role of light relic particles in the structure and history of the universe, tests of gravity at very large scales, measurements of the emergence of clusters of galaxies, time domain observations of transients at millimeter-wavelengths and even the exploration of our Solar System. The non-inflationary science goals of CMB-S4 drive the need for wide, arcminute-resolution observations of the millimeter-wave sky, while the inflationary science goal drives the need for a deep, degree-resolution survey. The inflation science goal also benefits from arcminute observations, which enable corrections of the B-mode signal in a processing step often referred to as delensing. Furthermore, the deep delensing survey benefits from a wide field of view with hundreds of thousands of detectors giving uniform frequency coverage on overlapping patches of sky \cite{abazajian2016cmbs4,abazajian2019cmbs4_projectplan}.

\setlength{\tabcolsep}{2.5pt}
\begin{table*}[!b] 
    \centering
    \begin{tabular}{rrrrrrrrrrrr}
        \toprule
         & $A_{0,1}$ & $A_{2,0}$ & $A_{0,2}$ & $A_{2,1}$ & $A_{0,3}$ & $A_{4,0}$ & $A_{2,2}$ & $A_{0,4}$ & $A_{4,1}$ & $A_{2,3}$ & $A_{0,5}$\\
        \midrule
        M1 & -4.9656 & -140.8171 & -116.1019 & 5.6312 & 4.1057 & 0.2358 & 0.0935 & -0.1069 & - & - & -\\
        M2 & -17.6056 & -403.0607 & -230.5055 & 61.6645 & 25.4229 & 11.6971 & -2.4272 & -3.5109 & - & - & -\\
        M3 & -22.1905 & -330.6599 & -280.4026 & 28.1685 & 17.4860 & -2.1208 & -10.8356 & -5.7779 & 0.8436 & 1.9139 & 0.6830\\
        \bottomrule\end{tabular}

    \caption{Polynomial coefficients ($A_{i,j}$) describing the three freeform mirrors (M1, M2, M3 for primary, secondary and tertiary) surfaces in their local  coordinate system, which is centered on each mirror (see Table \ref{tab:coordinate_defs}). These coefficients are used with Equation \ref{eq:surf} to fully describe the telescope mirror surfaces. Units are $\rm {mm}$, coefficients shown with a dash are zero. }
    \label{tab:surface_coeffs}
\end{table*}

Diffraction-limited large-aperture (6-meter class) telescopes are able to achieve arcminute angular resolution observations of the CMB at millimeter wavelengths. Next-generation high-sensitivity CMB observations require a large field of view to accommodate hundreds of thousands of detectors with minimal systematics. Unobstructed optical configurations provide low optical systematics, for which off-axis configurations have been used in many large-aperture CMB telescopes. In the past, off-axis Gregorian telescopes (where light focuses between the primary and secondary mirrors) \cite{Fowler:07, Padin:08} provided fields of view large enough to accommodate multiple cameras. Later the crossed Dragone  configuration  has been used \cite{2018SPIE10700E..41P} to accommodate a larger focal plane in CMB bands. Advances in technology have enabled the construction of receivers containing  increasing numbers of detectors, which are currently accommodated in crossed Dragone telescopes. The crossed Dragone configuration however produces rapidly changing astigmatism, which is hard to correct even with high order aspheric terms.

In the past, 6-meter class telescope mirrors have been manufactured by machining aluminum panels and carefully aligning them to form a large dish. Even with careful alignment, this approach leads to complex diffraction patterns from the small gaps between panels which spill power to large angular scales on the sky. These characteristics (astigmatism-related field of view limitations at $1.1\, \rm{mm}$ and panel gap diffraction) motivate the exploration of new technological solutions such as the incorporation of a third mirror, the use of freeform surfaces and the implementation of seamless mirrors to improve  performance in order to achieve the levels of sensitivity, wide field of view, uniform band coverage and low diffractive sidelobes required in the next generation of CMB experiments.

\begin{figure}[t]
    \centering
    \includegraphics[trim={2.7cm 0 1cm 0},clip, width=0.80\columnwidth]{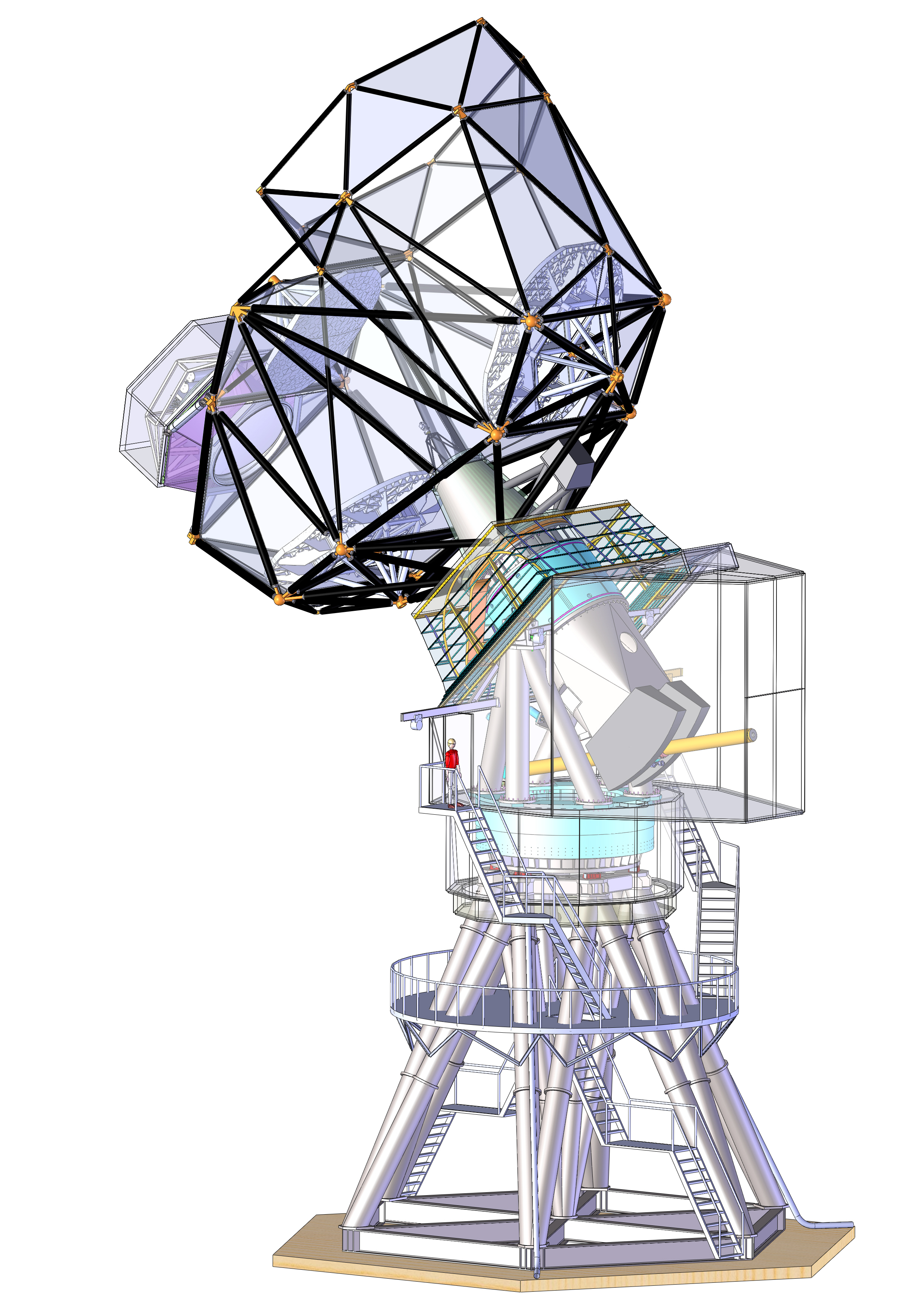}
    \caption{A rendering of the three-mirror anastigmatic concept for the South Pole CMB-S4 large-aperture telescope.}
    \label{fig:render}
\end{figure}

Three-mirror telescopes are able to cancel all first-order aberrations (including astigmatism), achieving a large  field of view with  diffraction-limited performance. Notable three-mirror telescopes include JWST and  future astronomical observatories such as the Vera C. Rubin Observatory or the E-ELT \cite{jwst.10.1117/12.559871, 2019ApJ...873..111I, 2007Msngr.127...11G}, however these optical and infrared telescopes are obstructed by their secondary mirror and its supporting structure, which introduces beam asymmetries. An unobstructed off-axis three-mirror telescope design concept for CMB observations has been proposed \cite{ApOpt..57.2314P} with standard surfaces (described by a radius of curvature, and a conic constant). This telescope concept has been designed to be manufacturable with 5-meter monolithic mirrors. Monolithic mirrors provide an attractive balance between manufacturability and a low level of large-angle diffraction. In this paper we explore such a three-mirror telescope and provide an updated design. This telescope design has 9.4 degrees of field of view with excellent image quality at $\lambda = 1.1 \, \rm mm$ and a lower f-number, which allows the same camera concept to be used as for the crossed Dragone design. This facilitates the cryomechanical design and minimizes development complexity. We provide an updated design for a set of 85 cameras (following up on previous work \cite{Gallardo2022:10.1117/12.2626876}) that populate the focal plane and show expected performance.

This article is part of a series of development studies to evaluate performance for the South Pole three-mirror large-aperture telescope for CMB-S4 including sidelobes performance \cite{sidelobes_Gullett:23} and seamless mirror manufacture \cite{fabrication_Natoli:23}.
This work presents a three-mirror polynomially defined freeform large-aperture (5 - meter) telescope and 85 three-lens cameras for CMB observations. This telescope design evolved from the three-mirror system presented in Padin et al. \cite{ApOpt..57.2314P} (herein P18) with a lower f-number (f/2.6 vs f/3.7) and it was optimized taking into account manufacturability constraints. The three-lens camera design is based on heritage technology from previous CMB experiments \cite{2016ApJS..227...21T,  10.1117/12.2312971, 2018SPIE10700E..3ED, 2021ApOpt..60..823G,ruhl04,Carlstrom_2011,Padin:08,Henderson_advactpol_2016JLTP..184..772H,George_etal_10.1117/12.925586,Sobrin_2022}  which minimizes engineering risks. The array of 85 cameras can be built using only two optical prescriptions for three silicon lenses as it was briefly described in \cite{Gallardo2022:10.1117/12.2626876}. This telescope design concept and its array of cameras is intended to be deployed as the South Pole large-aperture telescope for CMB-S4, a large-scale CMB observatory, which will provide maps of the cosmic microwave background with unprecedented levels of sensitivity and wide area coverage.

This article is structured as follows: Section \ref{sec:introduction} gives a summary of the state of the art in large-aperture telescopes for CMB bands and summarizes the aim of this study. In Section \ref{sec:tma_designdescription} we give a detailed description of the optics of the updated three-mirror anastigmatic telescope. Section \ref{sec:tel_nominalperformance} presents the nominal performance for the three-mirror system. Section \ref{sec:tel_tol} shows a tolerance analysis of this three-mirror system. Section \ref{sec:camera_design} describes the design of the array of cameras for the telescope. Section \ref{sec:camper} describes the performance of the camera system in conjunction with the three-mirror telescope optics. Section \ref{sec:tel_cam_tol} contains a tolerancing analysis for one single camera of the array, in conjunction with the three-mirror system. We conclude in Section  \ref{sec:conclusion}.

\section{Telescope  design description}
\label{sec:tma_designdescription}

A previous optical design of  a three-mirror telescope for CMB observations proposed an off-axis configuration where the tertiary is approximately the same size as the primary. The size of the tertiary is driven by the cancellation of  astigmatism over a wide field of view, which can be as large as 9 degrees at $1.1 \, \rm mm$ in this configuration. A system with these characteristics has been demonstrated in \padinetal, which follows work in three-mirror telescopes for on-axis configurations \cite{Korsch:77}. 
 
\begin{figure}
    \centering
    \includegraphics[width=0.70\columnwidth, trim={0.5cm 0 2.5cm 1.8cm}, clip]{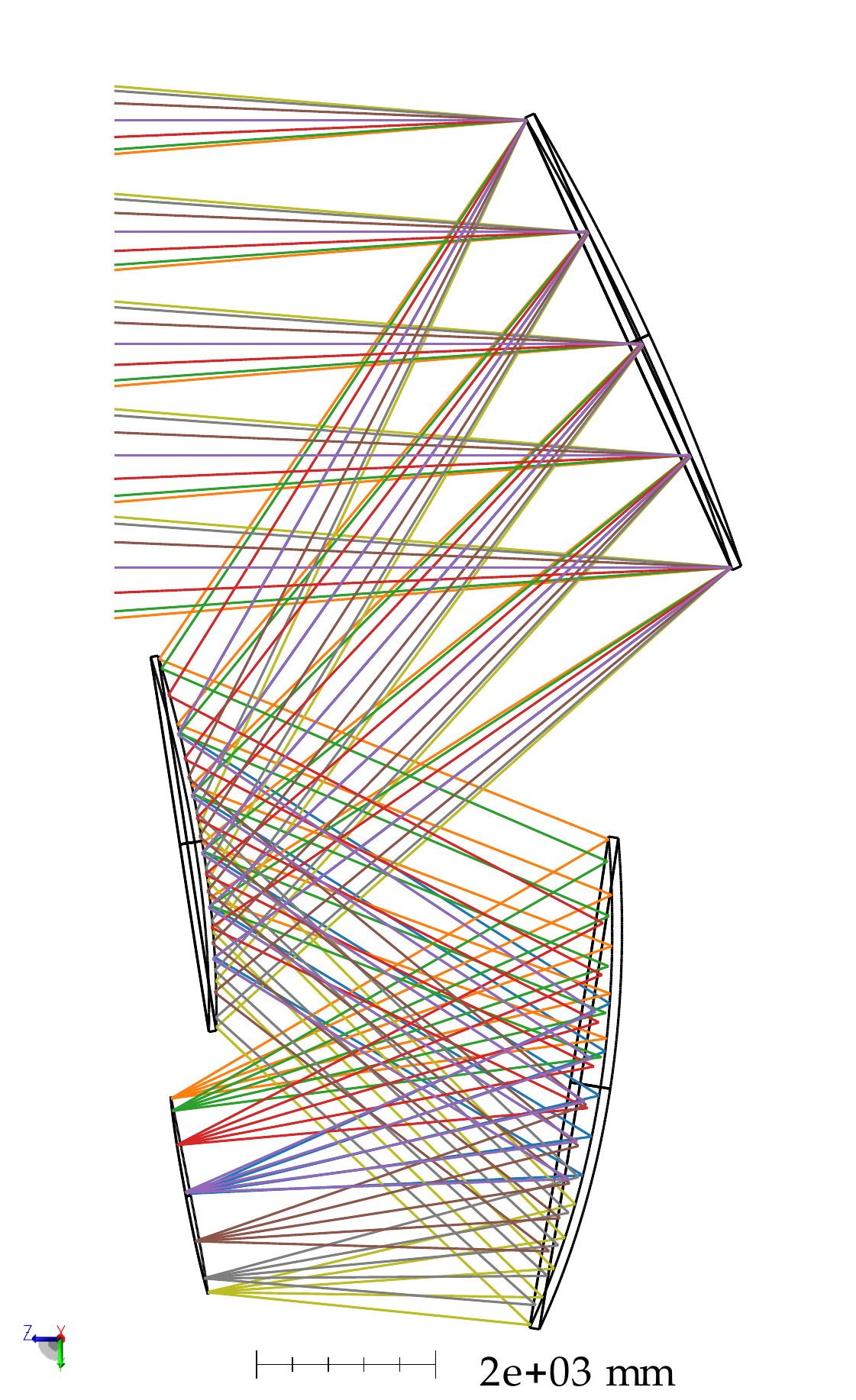}
    \includegraphics[width=0.93\columnwidth, trim={0.5cm 0.0cm 0.25cm 1cm}, clip]{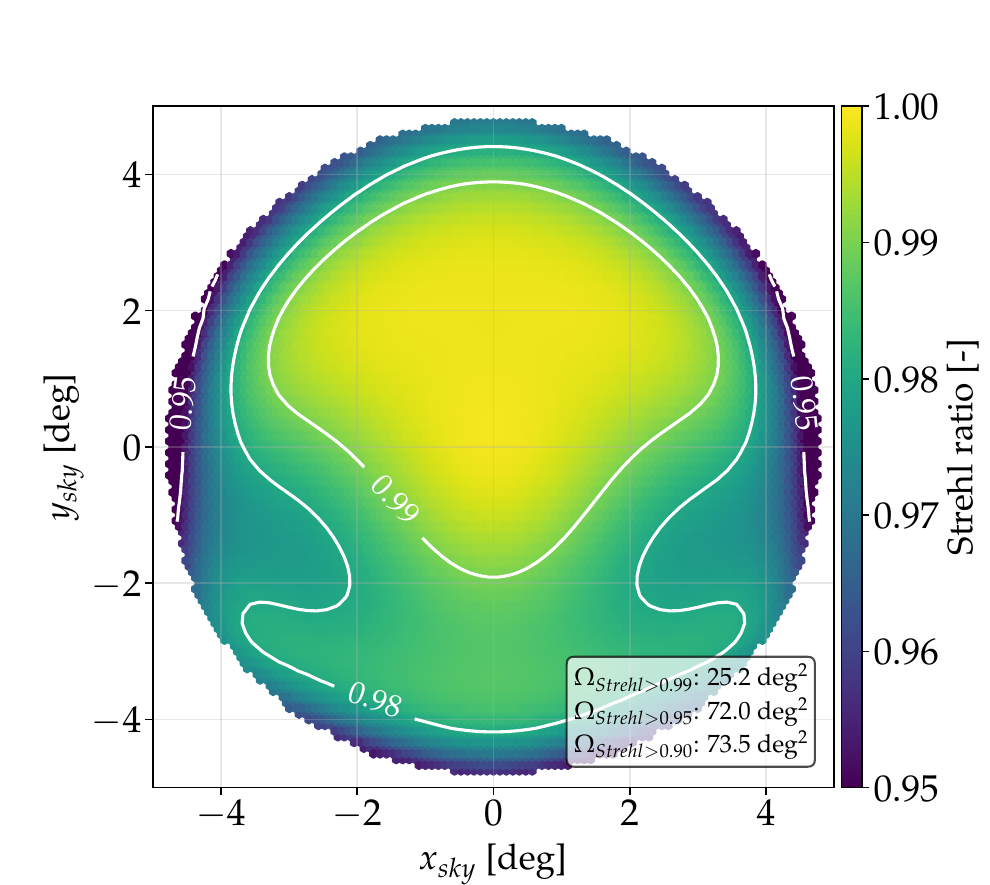}
    \caption{Top: Optical layout of the three-mirror anastigmatic telescope (TMA). Colored rays show extreme fields 4.7 degrees from the boresight. Bottom: Unvignetted Strehl ratios at $1.1\, \rm mm$  for a focal plane of 1.1 meters in radius and a field of view of 4.7 degrees in semi-diameter. The field of view in this system is limited by vignetting given by  the size of the tertiary. Contours and solid angle are shown for Strehls of 0.99, 0.98 and 0.95. \label{fig:layout_and_strehls_TMA}
    }
\end{figure}

\begin{table}[htb]
    \centering
    \begin{tabular}{lrrrr}
        \toprule
        Surface & X [mm] & Y [mm] & Z [mm] & $\alpha$ [deg] \\
        \midrule
             M1 &    0.0 &    0.0 &    0.0 &      155.40 \\
             M2 &    0.0 & 5615.0 & 4898.0 &      171.16 \\
             M3 &    0.0 & 8367.0 &  445.0 &     -170.88 \\
             FP &    0.0 & 9575.8 & 5024.4 &      168.93 \\
        \bottomrule
        \end{tabular}

    \caption{Positions and rotation angles describing the local coordinate systems of the three mirror surfaces (M1, M2 and M3) and the focal plane (FP). Origin of coordinates lies at the center of the primary mirror M1, z-axis points towards the boresight and the x-axis points into the page. Rotations shown are with respect to the x-axis and follow the right-hand convention (clockwise is positive).}
    \label{tab:coordinate_defs}
\end{table}

The design presented in {\padinetal}  used a concave primary mirror, a convex secondary and a concave tertiary. In the updated design, we maintain this optical configuration to allow compatibility with existing mechanical engineering development, and we modify the design to yield a lower f-number ($f/2.6$ vs $f/3.7$), which enables a smaller and lighter instrument, illuminated with high Strehl ratios at wavelengths as short as $1.1\, \rm mm$. This lower f-number is well-matched to the existing crossed Dragone design to be used in the CMB-S4 large-aperture telescopes planned to be deployed in Chile, which are also f/2.6 systems \cite{2018SPIE10700E..41P}. This lower f-number results in a tertiary mirror that is closer to the focal plane. In this design the primary and tertiary mirrors are of comparable size and within manufacturability constraints given by mechanical engineering considerations. Figure \ref{fig:layout_and_strehls_TMA} (top) shows the optical layout of this updated design.

The three reflective surfaces of the updated three-mirror telescope design are defined with two-dimensional freeform polynomial surfaces in rectangular coordinates. For each mirror, a local coordinate system is defined by translating the origin and rotating the coordinate system around the x-axis. The polynomially defined surface for each mirror is given by \begin{equation}
    z(x, y) = \sum_{i, j}^{i,j\in 0..5} A_{i,j} (x/R)^i(y/R)^j,
    \label{eq:surf}
\end{equation}
where $A_{i, j}$ is the coefficient for the term $x^iy^j$  and $R$ is a normalization radius (equal to 2500 mm). This parametrization has the property of normalizing the $x$ and $y$ coordinates, such that the coefficient $A_{i,j}$ is the magnitude of the sag (the deviation from flat) of the $x^iy^j$ polynomial term. Table \ref{tab:surface_coeffs} shows the coefficients that describe the mirror surfaces for the three mirrors. Rotation angles and locations for the local coordinate systems are shown in Table \ref{tab:coordinate_defs}. This telescope design yields a moderately curved focal plane which departs from a plane a maximum of $55\, \rm mm$  ($20\, \rm mm$) over $1.1$ meters in the x (y) direction. This field curvature can be corrected using alumina prisms, which flatten the field and allow the illumination of an array of cameras with parallel optical axes as described in Section \ref{sec:camera_design}.

\begin{figure}[htb]
\centering
\begin{tikzpicture}
    \draw (0, 0) node[inner sep=0] {\includegraphics[width=\columnwidth]{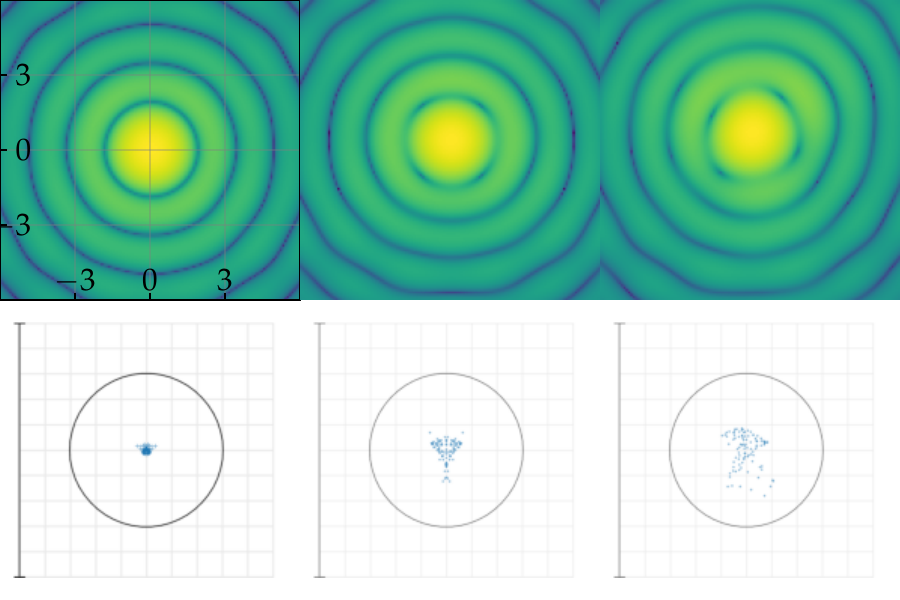}};
    %\draw (-4.35, -1.50) node {\rotatebox{90} {3 arcmin}};
    \draw ( -1.45, -1.50) node {\rotatebox{90} {3 arcmin}};
    %\draw (1.5, -1.50) node {\rotatebox{90} {3 arcmin}};
\end{tikzpicture}
\caption{Three-mirror telescope beam shapes and spot diagrams in sky coordinates (units are arcminutes).  Top: Monochromatic beam shapes at $\lambda = 1.1 \, \rm{mm}$ computed with a Huygens diffraction model (Zemax) using uniform illumination of the stop and an aperture of 5 meters in the time forward direction.
Bottom: Corresponding spot diagram for the center and two extreme field positions 4.7 degrees from the boresight. Field positions in the focal plane are (from left to right) (0, 0) mm, (0, 1100) mm and (1100, 0) mm. All rays lie within an Airy disk radius given by $1.22\frac{\lambda}{D}$ for a wavelength of $1.1\, \rm mm$ and a diameter of 5 meters. Strehl ratios for these three fields are 0.99, 0.97 and 0.95.}
\label{fig:spot_diag_and_beams}

\end{figure}

\section{Telescope Nominal Performance}
\label{sec:tel_nominalperformance}

Image quality for the three-mirror telescope was quantified using Zemax OpticStudio at $\lambda = 1.1 \, \rm mm$ in the time-forward direction (sky $\rightarrow$ mirrors $\rightarrow$ focal plane) using a circular aperture of 5 meters perpendicular to the boresight. The telescope yields Strehl ratios above 0.94 over $1.1\, \rm meters$ or 4.7 degrees in semi-diameter. Figure \ref{fig:layout_and_strehls_TMA} (bottom) shows  Strehl ratios over the unvignetted field of view. Figure \ref{fig:spot_diag_and_beams} shows simulated Huygens diffraction beams and spot diagrams for this system.

\begin{figure}
    \centering
    \begin{subfigure}[t]{\columnwidth}
    	\includegraphics[width=\textwidth]{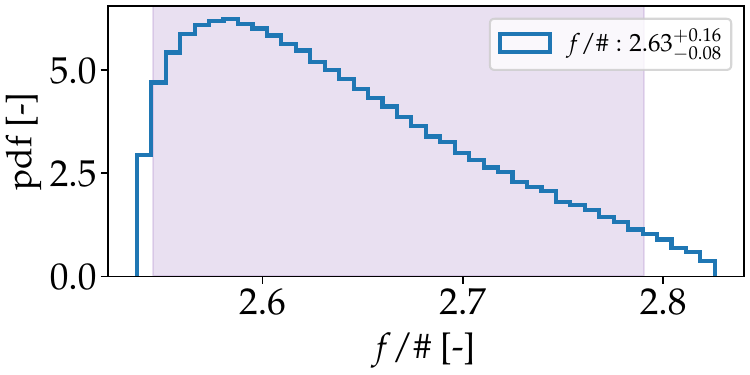}
    \end{subfigure}
    \caption{Chief ray-relative average f-number for the three-mirror telescope across the x-y directions. Label denotes the median f-number with $95\%$ limits across all fields in the sky.}
    \label{fig:fnum}
\end{figure}

We quantify f-numbers by tracing rays in the time-forward direction and computing the angle formed by the chief ray and the marginal ray in four directions at the focal plane. We compute the f-number according to $f/\# =1/2\tan(\theta)$. Figure \ref{fig:fnum} shows a histogram of the chief ray-relative f-number, which has a median value of 2.63. As shown in Figure \ref{fig:layout_and_strehls_TMA} (top) and discussed in Section \ref{sec:tma_designdescription}, the focal surface is moderately curved. The angle between the chief ray and the focal surface in this design stays between zero and 8 degrees. This level of field curvature can be corrected using a prism (with a unique tilt and clocking for each camera) as discussed in Section \ref{sec:camera_design}. The cone of light reaching the focal plane (for a circular stop in front of the primary mirror) is circular (to within 1\%) at the center of the field of view and has a varying ellipticity across the focal plane, with a maximum ellipticity of 9\% at the bottom.

\section{Telescope Tolerances}
\label{sec:tel_tol}
Tolerance analysis of a freeform system is complex due to the many non-orthogonal parameters that define the optical surfaces and the correlations among them. In addition, the presence of gravitational and thermal deformations further complicate the analysis. To simplify the tolerance analysis of the three-mirror telescope, we split it as follows: 1) first we compute a classical tolerance analysis, where we vary the positions and tilts for the three mirrors in the system one by one and jointly in a Monte Carlo simulation (with and without refocusing) and 2) we implement gravitational and thermal deformations obtained from a computational finite element analysis model to estimate the degradation in image quality due to the deformation of the mirrors.

\subsection{Tolerancing in mirror placement and tilts}
\label{sec:tolplacements}

We use a three-mirror telescope time-forward model, where we vary the distances between mirrors (3 variables), tilts (6 variables) and mirror decenterings (6 variables). We define a merit function as the minimum Strehl ratio at 8 positions located 4.7 degrees from the boresight. The minimum Strehl ratio is used to avoid averaging over a large area with uniform Strehl ratio which biases the average high. This metric is more stringent than simply taking the average Strehl ratio. We find values for these 15 variables that degrade this merit function by 0.05. This degradation is found by varying these 15 parameters one at a time, while keeping the rest of the parameters in their nominal values. We find that for this criterion, an inter-mirror distance variation of $\pm5 \, \rm mm$, a mirror decentering of $\pm5 \, \rm mm$ and $\pm0.09$ degrees of tilt results in a five percent degradation without refocusing compensation. Refocusing compensation allows a relaxation in the tolerance for mirror placement to $\pm 14\,\rm mm$ while keeping the tolerances in decentering and tilts unmodified. These values are summarized in Table  \ref{tab:tel_tols}, where columns labeled \emph{Tol.} and \emph{Tol. Ref.} show the single variable tolerances without and with refocusing compensation, respectively.

\begin{table}[t]
    \centering
        \begin{tabular}{ccccc}
        \toprule
        Parameter &  Tol. & Tol. MC & Tol. Ref. & Tol. Ref. MC\\
        \midrule
        Mirror dist. [mm] &       5    &    2   &     14   &  6  \\
        Decenter [mm]     &       5    &    2   &      5   & 2   \\
        Tilt [deg]        &    0.09    & 0.03   &   0.09   & 0.03\\
        \bottomrule
        \end{tabular}
    \caption{Tolerances for the LAT telescope without cameras. Tolerances for single parameter variations produce a degradation in  Strehl of $5\%$. Single parameter tolerances are shown without refocusing (column Tol.) and with refocusing (column Tol. Ref.). Tighter parameters are obtained with a joint Monte Carlo model, where all parameters are randomly varied following a normal distribution. Monte Carlo runs are shown without refocusing and with refocusing (Tol. MC and Tol. Ref. MC respectively). Note that a 0.03 degree tilt in a $5 \, \rm m$ mirror-yields a $2.7 \, \rm mm$ peak-to-peak surface error.}
    \label{tab:tel_tols}    
    \end{table}

In a Monte Carlo simulation we jointly vary the 15 variables reducing the maximum allowable variation found in the previous paragraph. The maximum allowable variation is obtained by scaling the individual variation  by a factor of $1/2.4$. This keeps the  Strehl ratio degradation approximately in the same scale as our 5 percent limit under assumption of uncorrelated random variations (assuming independence and linearity, the merit function scales approximately with $1/\sqrt{15}$). We find that a degradation of less than 5 percent in the merit function happens $99\%$ of the time for tolerances better than $2 \, \rm mm$ in inter-mirror distances, $2\, \rm mm$ in decenters and 0.03 degrees in mirror tilts. Refocusing improves the tolerance in mirror distances to $6\, \rm mm$, while keeping the decenter and tilt tolerances unmodified. The result of this Monte Carlo simulation can be interpreted as the worst case due to the unexplored correlations among variables, which are expected to reduce independence among the random variables in the simulation.

\subsection{Gravitational Deformation}

Gravitational deformations are predicted using a finite element analysis model of the primary and tertiary mirrors. This mechanical model includes the mirror's backing structure with the appropriate support mounting points. The deformation for this surface shows a peak next to a valley on the y direction with a root-mean-square (RMS) value of $10 \, \mu \rm m$ for the primary and $20 \, \mu \rm m$ for the tertiary. We model the secondary with a scaled down version of the deformations of the primary in the opposite direction because the secondary is convex.

We fit a polynomial $f(x, y)$ to the deformation surface in the same format as Equation \ref{eq:surf}. This polynomial form is convenient for optical modeling  as the perturbed mirror shape can be found by straightforward addition. We find that a polynomial of fifth degree adequately fits this surface with a residual RMS lower than  $3 \,  \mu m$ for the primary mirror. This three-mirror deformation results in a displacement of the telescope focal point by $1.4 \, \rm mm$ away from the tertiary mirror, decreasing the lowest Strehl ratio (which corresponds to the top side of the focal plane) by 2\%, i.e. to 0.94.

\subsection{Thermal Deformation}

A thermal model of the primary mirror is used to estimate the cupping due to the differential thermal contraction between the mirror backing structure and the optical side of the mirror. We obtain a sag of $60\, \mu \rm {m}$ for the primary mirror. We scale the sag of the primary mirror with the diameter of the secondary and tertiary mirrors to estimate their sag assuming a linear relation with mirror diameter to obtain sags of 39 and 60 $\, \mu \rm {m}$ respectively. We convert the sag of the perturbation to a radius of curvature using the equation \begin{equation}
R = \frac{r^2 + \delta z^2}{2\delta z},
\end{equation} where $R$ is the radius of curvature of the perturbing surface, $r$ is the radius of the mirror aperture and $\delta z$ is the sag of the perturbation obtaining radii of curvature of $5.33\times 10^7\rm{mm}$ for the primary and tertiary and $3.50\times 10^7 \rm{mm}$ for the secondary mirror. This deformation moves the focal plane forward (towards the tertiary) $9.7 \rm{mm} $, which can be refocused obtaining similar performance to the unperturbed case.

\begin{table}
\begin{center}
\begin{tabular}{c c c c} 
     \hline
     Deformation & Min Strehl [-] & Max Strehl [-] & Defocus [mm] \\
     \hline\hline
    Nominal & 0.96 & 0.99 & 0 \\ 
    \hline
    Thermal & 0.96 & 0.99 & -9.69 \\
    \hline
    Gravity & 0.94 & 0.99 & -1.36 \\
    \hline
    Thermal + Grav. & 0.92 & 0.99 & -10.24 \\
 \hline
\end{tabular}
\end{center}
\caption{Telescope Strehl ratio variations due to mirror deformations. The  image quality at the center field stays without variation, while the mirror shape distortions degrade the top field image quality by 4\%.}
\label{tab:teldefthermgrav}
\end{table}

\subsection{Thermal + Gravitational deformation}

We combine the thermal and gravitational deformation mentioned above and refocus the system. The new focal point is located $10.24\, \rm{mm}$ away from the unperturbed focal surface, we find that the minimum Strehl for this configuration is 0.92, while the maximum Strehl ratio stays at 0.99, showing that the combined deformation only degrades the minimum Strehl ratio by roughly 4\% if refocusing is allowed. Table \ref{tab:teldefthermgrav} shows a summary of the nominal, thermal and gravitational deformation Strehl ratios and their refocus amplitude.

\section{Camera design}
\label{sec:camera_design}
 
\begin{table}[t]
    \centering
    \begin{tabular}{llllllllll}
        \toprule
         $R_1$ & $R_2$ &  $R_3$ & $k_1$ & $k_2$ & $k_3$ & $th_1$ & $th_2$ & $th_3$ & $th_4$ \\
        \midrule
        -432.8 & 467.4 & -899.4 &  -9.9 &  -7.9 & -10.7 &  120.0 &  143.4 &   67.3 &  236.4 \\
        -422.9 & 458.3 & -923.6 & -10.3 &  -8.0 &  -9.9 &  120.6 &  139.3 &   66.4 &  246.0 \\
        \bottomrule
        \end{tabular}
    \caption{Parameters  found with the optimization procedure described in Section \ref{sec:camera_design}. Distances between L1-L2, L2-Stop, Stop-L3 and L3-FP are denoted with $th_1$, $th_2$, $th_3$, $th_4$. All distances and radii of curvature $R_j$ have units of $\rm mm$.}
    \label{tab:camera_prescription}
\end{table}

\newcommand{\toppanelheight}{9cm}

\begin{figure*}[p!]
\centering
    \begin{subfigure}[b]{0.35\textwidth}
    		\centering
            \begin{tikzpicture}
    \draw (0, 0) node[inner sep=0] {\includegraphics[width= \toppanelheight, trim={0cm, 0.1cm, 0cm, 0cm}, clip, angle=90]{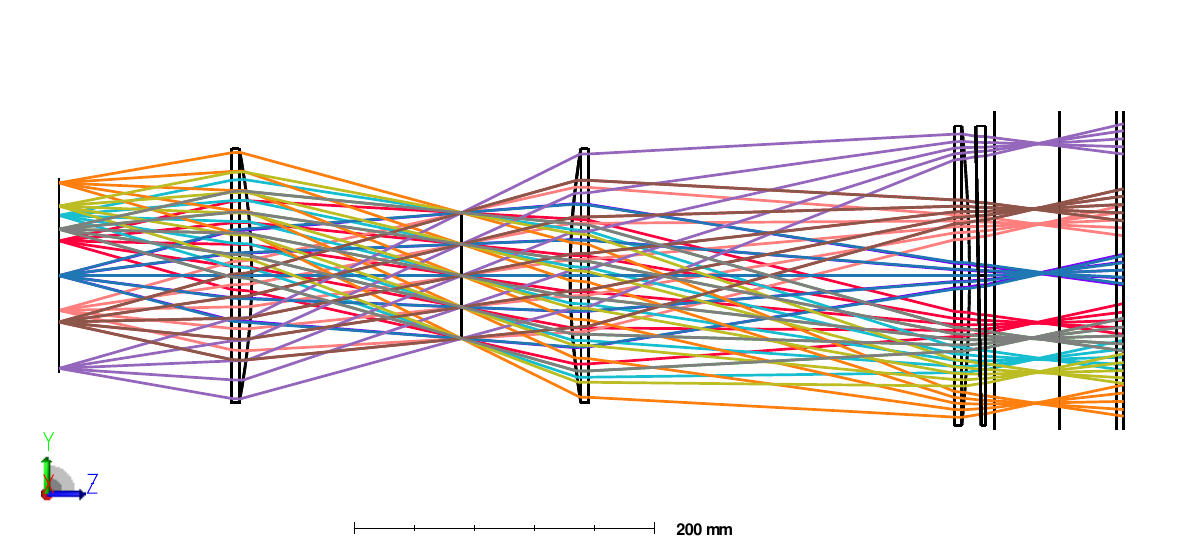}};
    \draw (1.75,  2.95) node {Wedge};
    \draw (-1.65,  2.75) node {L1 (4K)};
    \draw (-1.65, -0.05) node {L2 (4K)};
    \draw (1.3, -0.95) node {Lyot (1K)};
    \draw (-1.65, -2.7) node {L3 (1K)};
    \draw (-1.60, -3.9) node {FP (0.1K)};
\end{tikzpicture}
    \end{subfigure} 
    \begin{subfigure}[b]{0.20\textwidth}
    		\centering
            \begin{tikzpicture}
    		\draw(0, 0) node[inner sep=0]{\includegraphics[height=\toppanelheight, trim={2.5cm 0.2cm 2cm 3cm}, clip]{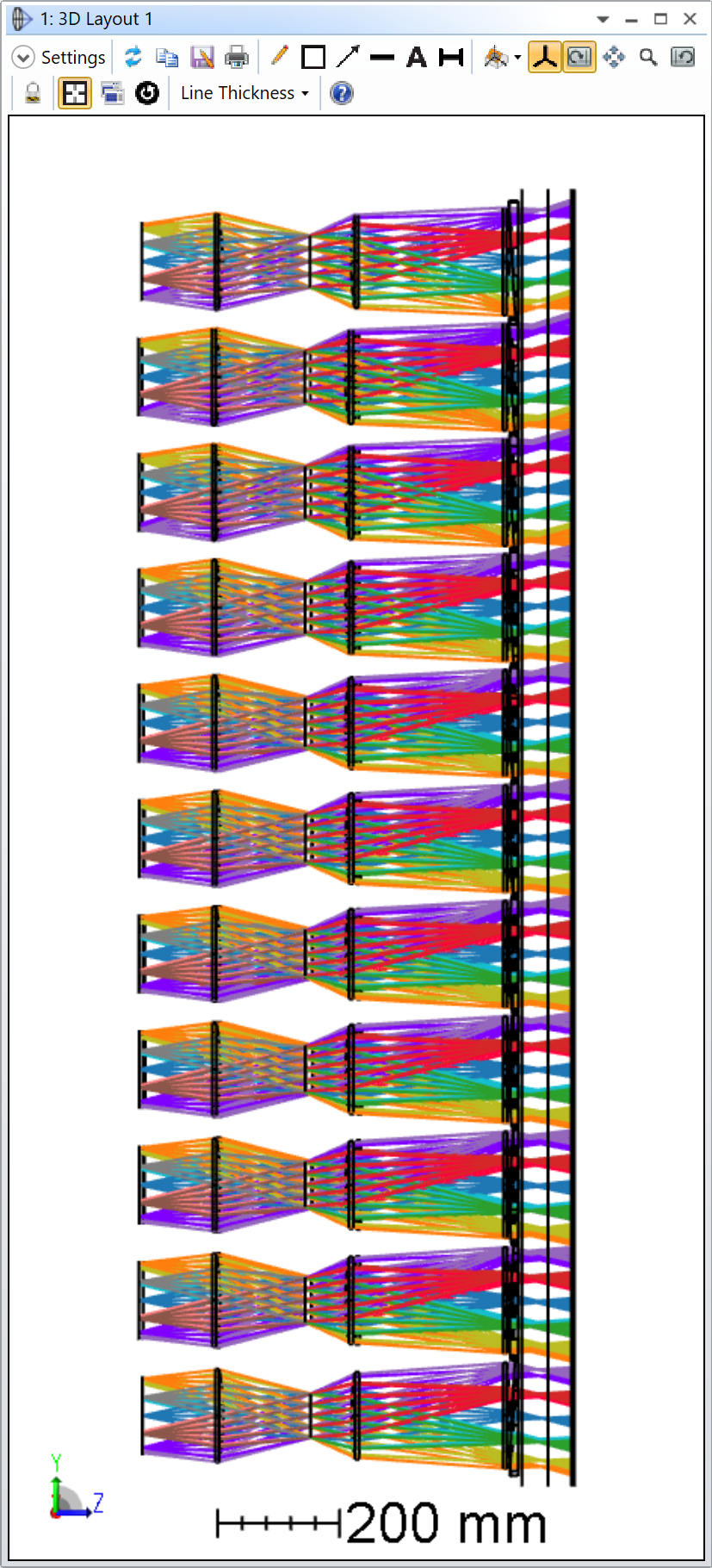}};
            \draw[-stealth](-1,4.45) -- (1,4.45);
            \draw[align=center] (0.0, 4.65) node {To Tertiary};
            \end{tikzpicture}
    \end{subfigure}
    \begin{subfigure}[b]{0.4\textwidth}
        \centering
            \begin{tikzpicture}
    \draw (0, 0) node[inner sep=0] {\includegraphics[height=\toppanelheight, trim={1.0cm 0.2cm 1cm 3cm}, clip]{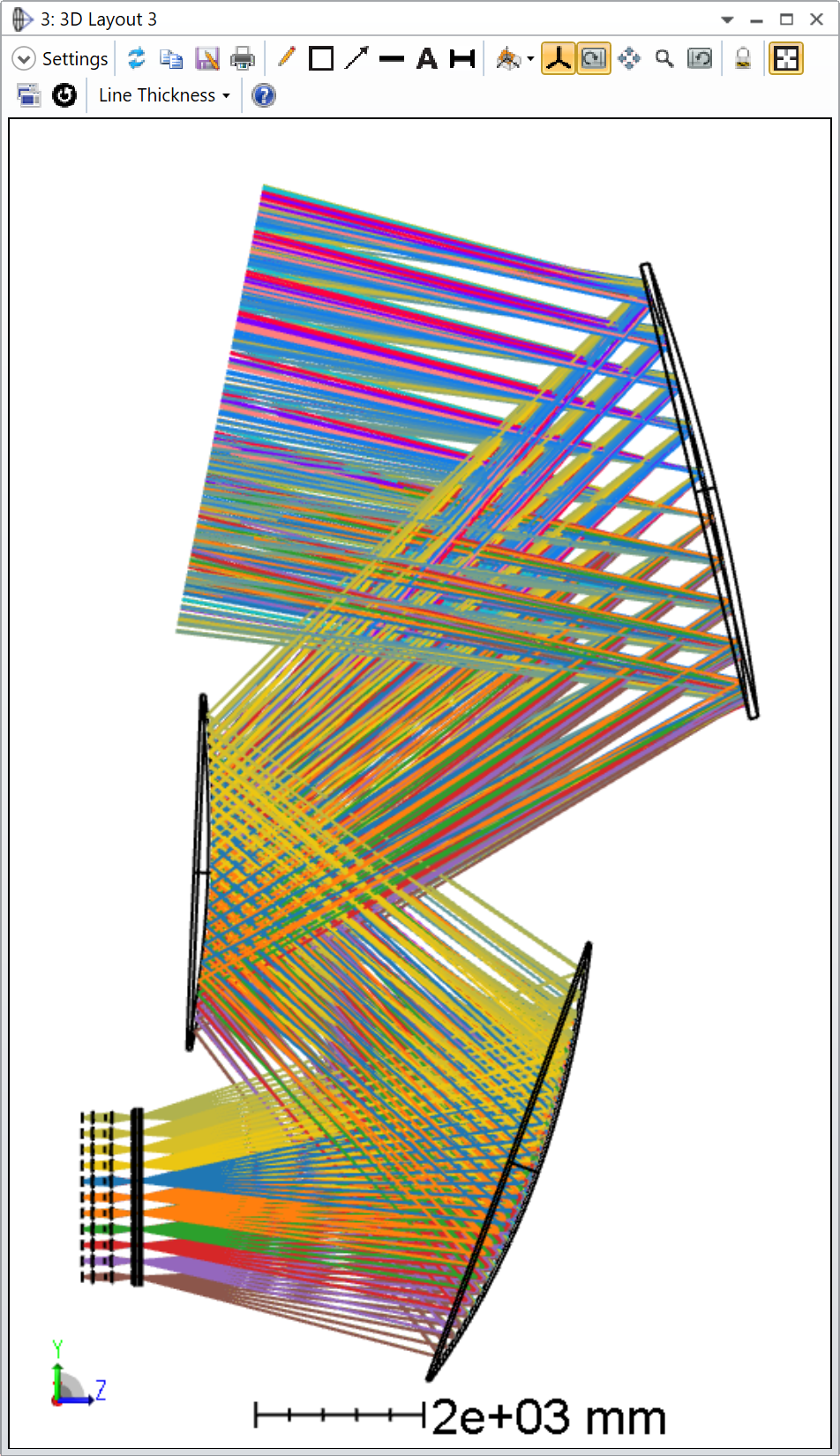}};
    \draw (2.80,  2.35) node {Primary};
    \draw (-2.3,  2.85) node {From Sky};
    \draw (-2.40, -0.4) node {Secondary};
    \draw (1.5, -2.7) node {Tertiary};
    \draw (-2.0, -3.7) node {Camera array};
\end{tikzpicture}
    		
    \end{subfigure}

\centering
    \begin{subfigure}[c]{0.45\textwidth}
        \includegraphics[width=\textwidth, trim={2.5cm 0.2cm 2cm 2cm}, clip]{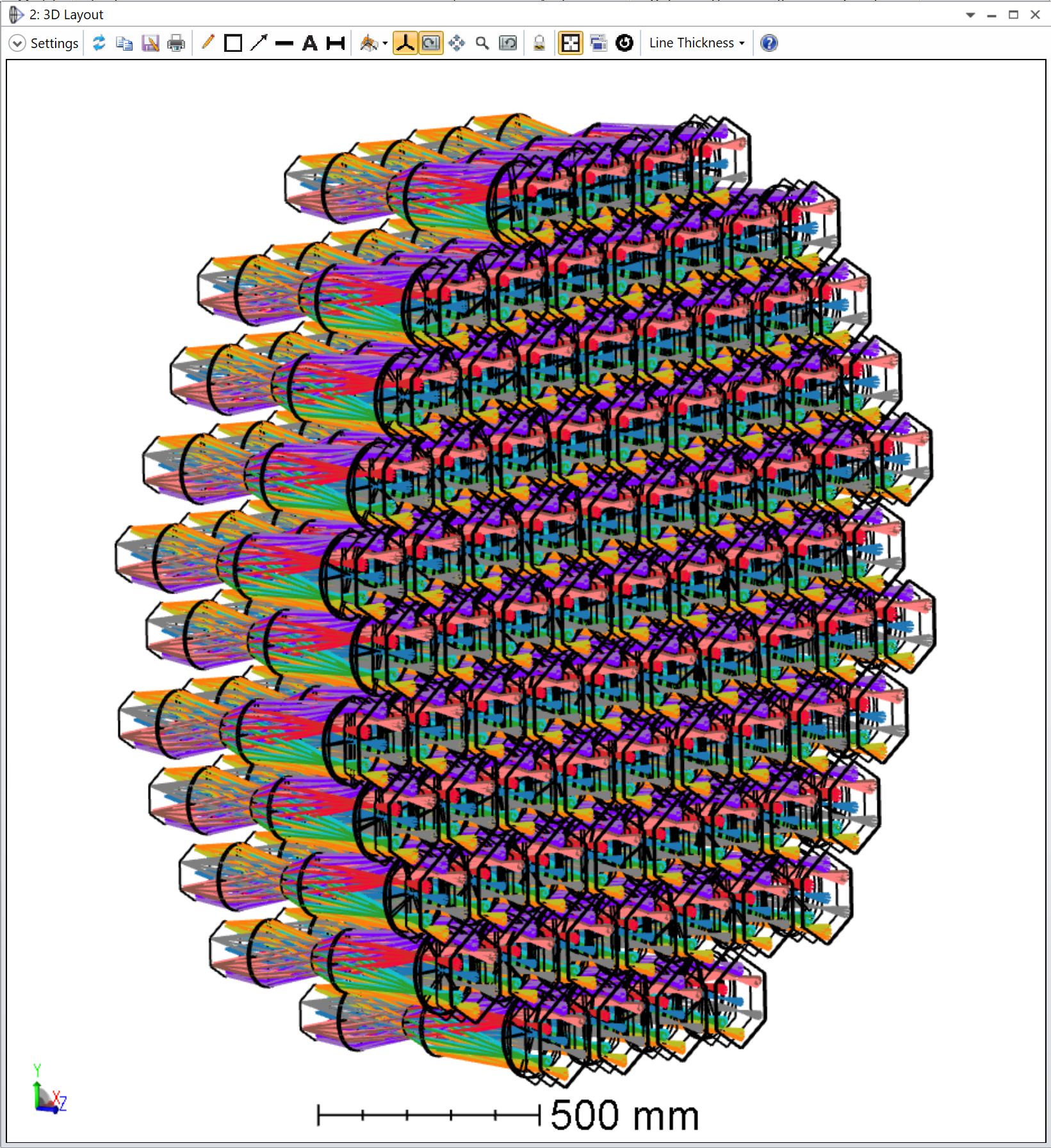}
    \end{subfigure}
    	\begin{subfigure}[c]{0.46\textwidth}
    \centering
    \includegraphics[width=\textwidth]{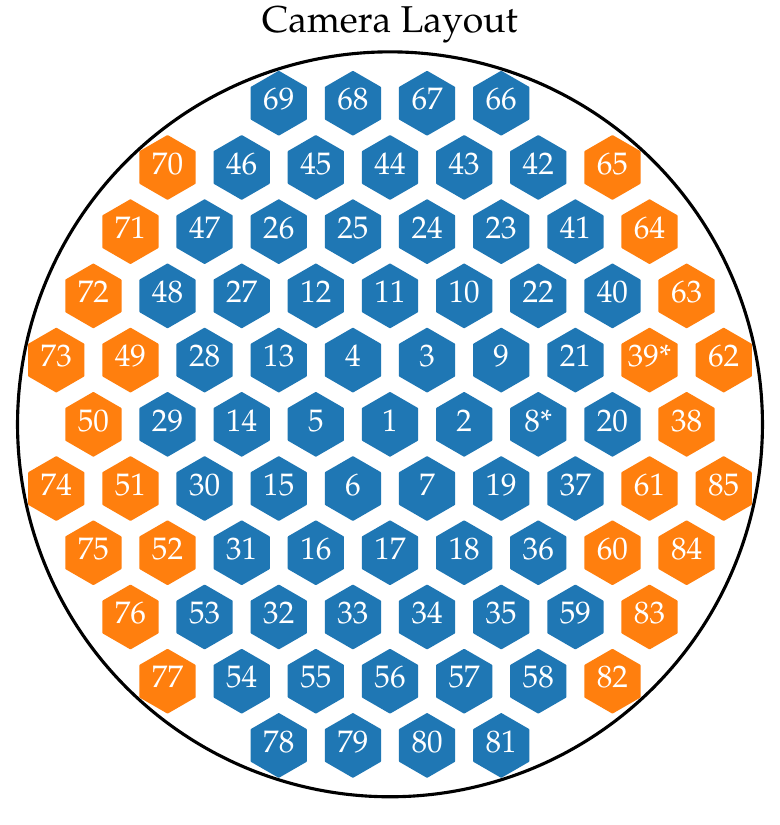}
    \end{subfigure}
    
    \caption{Top left: Optical layout for one of the 85 cameras (camera labeled 1) in the array, three silicon lenses named L1, L2, L3 (from top to bottom) and a Lyot stop are shown. A silica prism is placed in front of L1. Top center: side view of the 85 cameras and their prisms. Top right: side view of the three-mirror telescope with its 85 cameras. Bottom left: 85 camera array layout. Bottom right: camera arrangement and numbering convention used in this design. Cameras 8 and 39 have been optimized with the procedure described in section \ref{sec:camera_design}. Circle has a radius of 1100 mm.}
    \label{fig:camera-layouts}
\end{figure*}
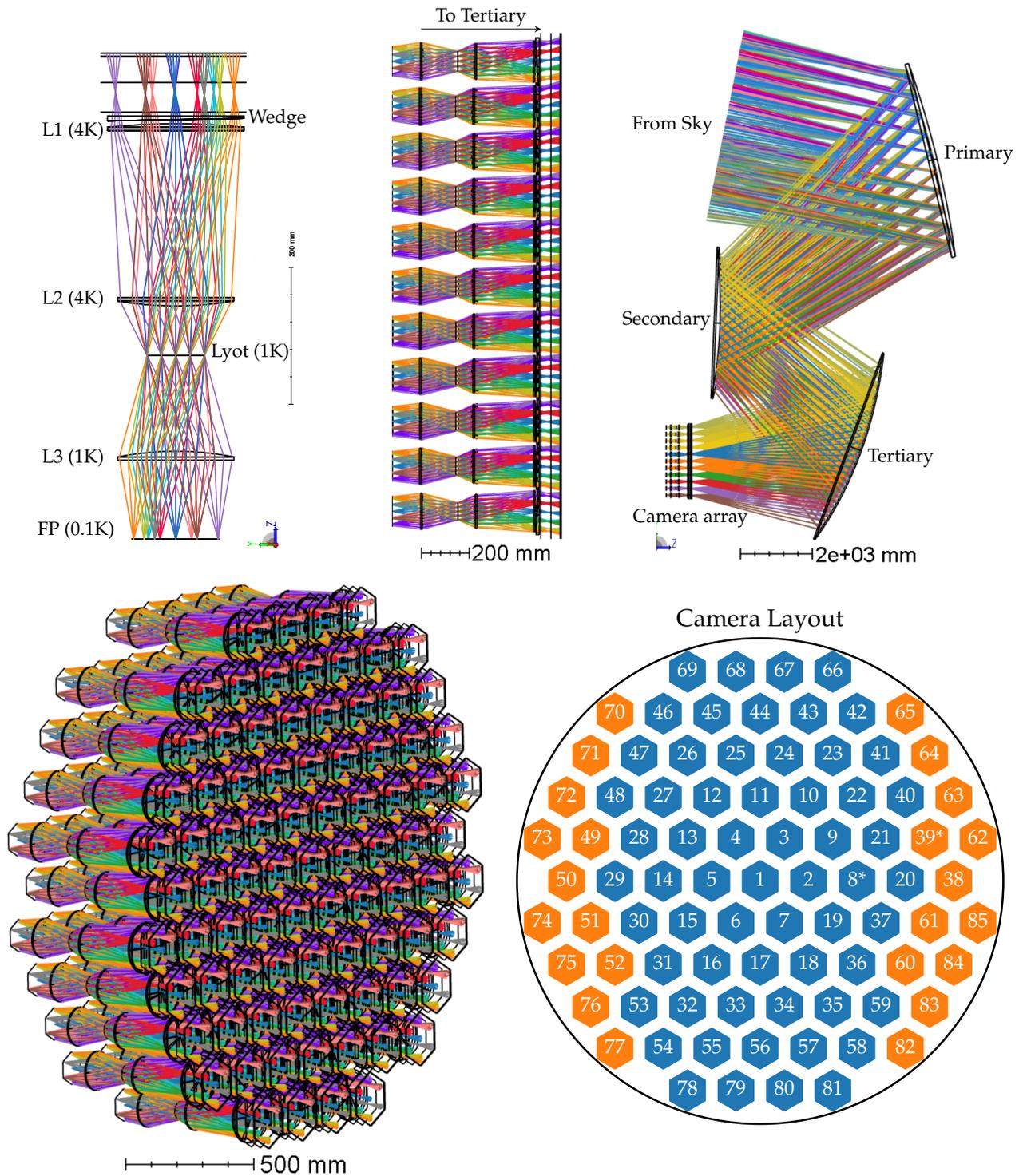

The receiver consists of an array of 85 three-lens cameras arranged in a hexagonal lattice following center positions $(x, y)$ given by 
 \begin{equation}
 \begin{bmatrix}x\\
y
\end{bmatrix} =
s \begin{bmatrix}
\sqrt{3}& \frac{\sqrt 3}{2}\\
0 & \frac{3}{2} 
\end{bmatrix}
\begin{bmatrix}
a\\
b
\end{bmatrix},
\end{equation} where $s=219/\sqrt{3}\, \rm{mm}$ is a scale parameter and $a$, $b$ are integer indices. Lenses in each camera are named L1, L2, L3 (ordered from telescope focal plane towards camera focal plane). A Lyot stop is placed between L2 and L3, which defines the primary mirror illumination.
Cameras have parallel axes and the first lens of all cameras (L1) is placed with its vertex located in the same plane.

All lenses are radially symmetric plano-convex, with the shape of the curved side given by \begin{equation}
    z = \frac{c r^2}{1 + \sqrt{1-(1+k)c^2r^2}},
    \label{eq:standard_surf}
\end{equation} where $c$ is the inverse of the radius of curvature $R$, $k$ is the conic constant and linear units are $\,\rm mm$. Field curvature from the telescope is corrected by an alumina prism with a tilt  and clocking that is found numerically in the time-reverse sense to center the footprint envelope for each camera on the primary mirror and to leave a clearance of $150 \, \rm mm$ from the primary rim, which helps shield against spillover.

Optimization of the cameras is performed in the time-reverse sense, with light rays starting at the focal plane of detectors (which has a diameter of 130 mm) passing through the three lenses and mirrors towards the sky. This time-reversed model allows a simple pupil definition, which directly controls illumination of the primary mirror. Optimization is started with approximate shapes for the three lenses (see \cite{Gallardo2022:10.1117/12.2626876} for more detail in the initial parameters) and the alumina wedge (which can be approximately obtained by pointing the chief ray towards the center of the primary), while keeping a fixed diameter for the Lyot stop of $43 \, \rm{mm}$ which was found to provide a camera length of about $600 \, \rm mm$ and provides reasonable lens curvature. A merit function is defined such that it minimizes the RMS spot-diameter size on the sky for fields distributed across the camera focal plane (see \cite{Gallardo2022:10.1117/12.2626876} for more details). We also include constraints that keep all rays within the diameter of the lenses and constraints that make a circular stop fully illuminated. The primary illumination is controlled approximately by sampling the marginal rays in the x-direction to roughly fill the primary. We optimize cameras over ten variables (6 variables controlling the shapes of the lenses according to Equation \ref{eq:standard_surf} and four variables controlling their spacings).  We optimize all 85 cameras individually and evaluate each camera to find if a solution can be successfully replicated to minimize manufacturing complexity. We find that two solutions for the shapes and distances of the lenses are enough to give an acceptable Strehl ratio across the field of view of the telescope. These two solutions cover: 1) the center region (61 cameras) of the array of cameras and 2) the two areas at the sides of the array (composed of 24 cameras) as shown in Figure \ref{fig:camera-layouts} (bottom right), where blue and orange denote the two blocks of cameras sharing the same prescription. Values for these two groups of cameras are shown in Table \ref{tab:camera_prescription}.

After finding the optimum values for the lens shapes and distances, the stop size is adjusted for each camera by tracing a ring of marginal rays onto the primary mirror and calculating the distance between this ring of rays and the mirror rim. The stop size is found by making this distance equal to $150\, \rm{mm}$ (which effectively uses part of the primary to control spilled power) while keeping the center of mass for this ring of rays centered at the primary origin. We do this for a circular stop over 3 variables (the stop size, camera wedge tilt and clocking angle), and an elliptical stop over 5 variables (stop semiaxes and clocking, camera wedge tilt and wedge clocking angle).  We evaluate the f-number as described in Section \ref{sec:camper} and conclude that between these two candidate designs the elliptical stop gives better primary mirror illumination and lower f-numbers  as shown in Figure \ref{fig:imgqual_fnums_prism} (bottom left).

\section{Camera performance}
\label{sec:camper}

\begin{figure}[htb]
    \centering
    \includegraphics[width=\columnwidth]{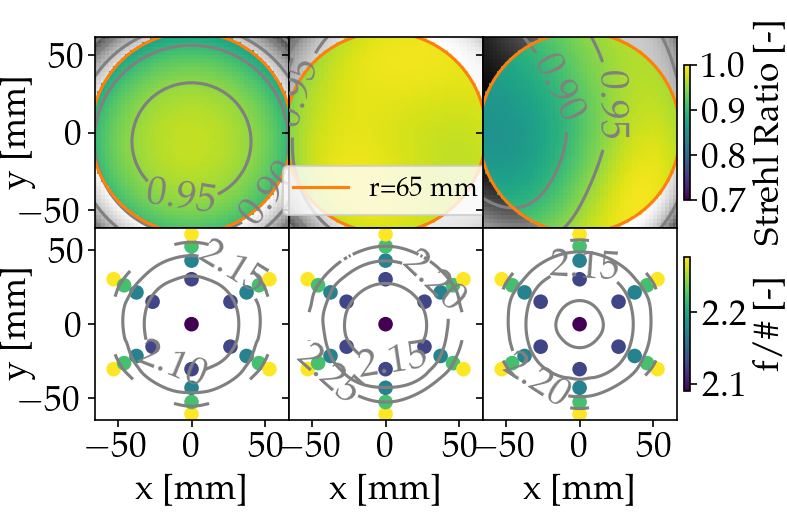}
    \caption{Strehl ratios at $\lambda=1.1\, \rm mm$ (top) and f-numbers (bottom) for  cameras 1, 31 and 83 (from left to right). Axes show positions at the detector focal plane.}
    \label{fig:strehls_fnumbs}
\end{figure}

\begin{figure*}
	\centering
	\begin{subfigure}[t]{0.48\textwidth}
	\includegraphics[trim= 0.35in 0.25in 0.35in 0.25in, clip, width=\textwidth]{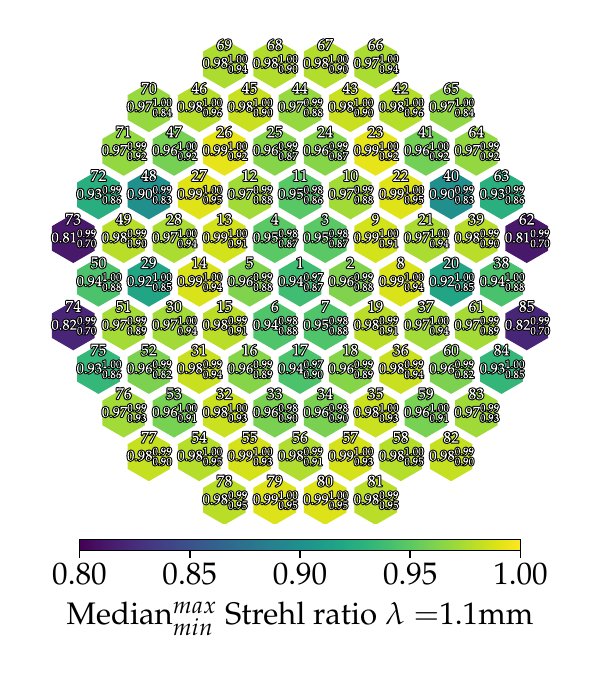}
	\end{subfigure}
    \begin{subfigure}[t]{0.48\textwidth}
    \includegraphics[trim=  0.35in 0.25in 0.35in 0.25in, clip, width=\textwidth]{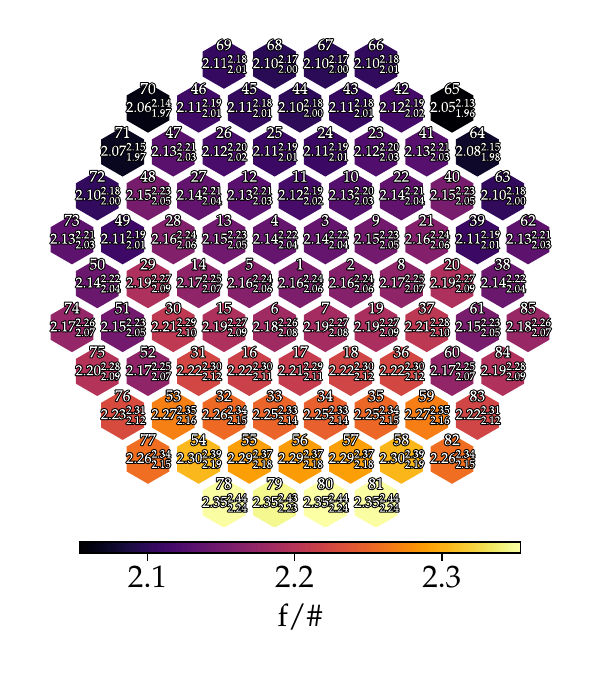}
    \end{subfigure}
 
	\centering
    \begin{subfigure}[t]{0.48\textwidth}
    \includegraphics[width=\textwidth]{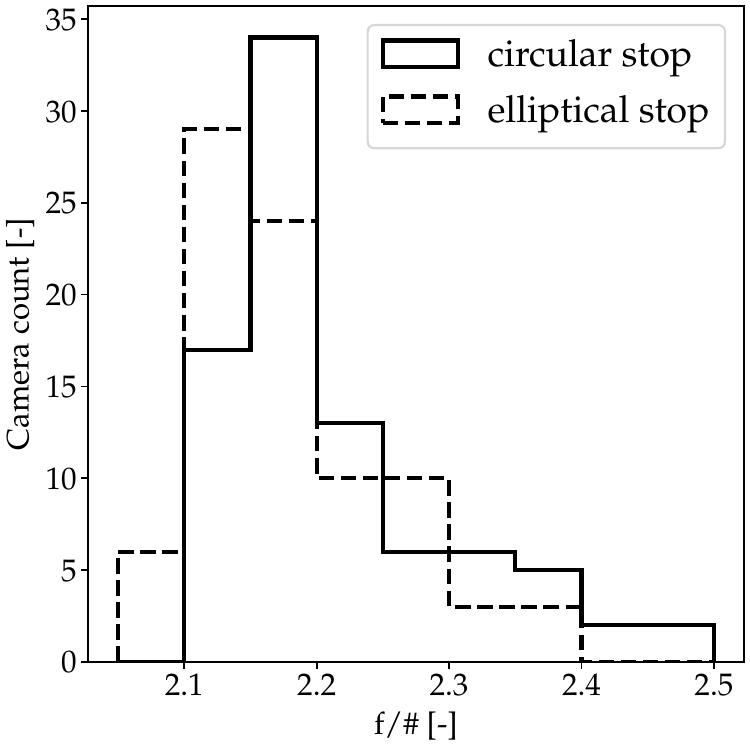}
    \end{subfigure} 
	\begin{subfigure}[t]{0.48\textwidth}
	\includegraphics[width=\textwidth, trim= 0.24in 0 0.22in 0, clip]{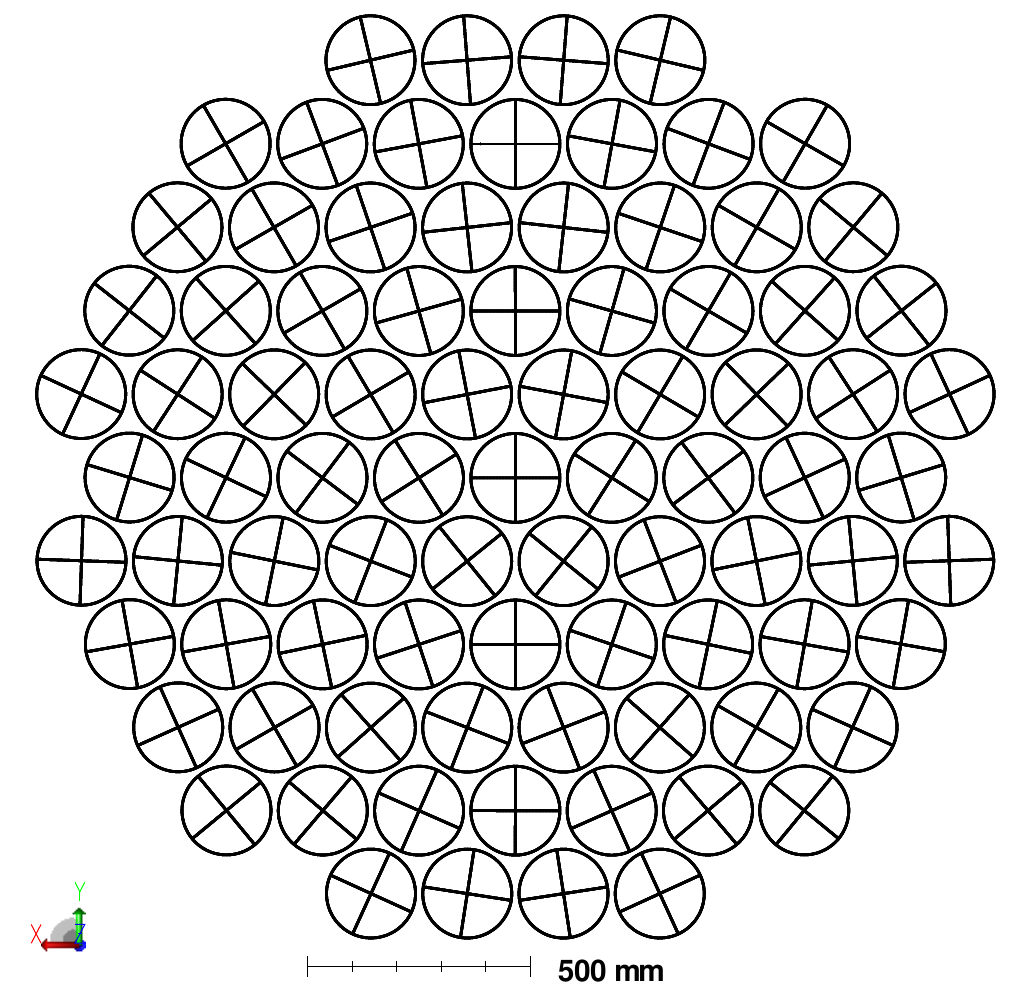}
	\end{subfigure}
    \caption{Top left: Median, minimum and maximum Strehl ratios for each of the 85 cameras at $\lambda=1.1\, \rm mm$.  
    Top right: f-number distribution across 85 cameras. Superscript (subscript) denotes the maximum (minimum) f-number value in a particular camera.
    Bottom left: histogram showing the average f-number distribution for all 85 cameras.
    Bottom right: alumina  wedge clocking for all cameras. Cross is shown as a comparison against the un-rotated x-y axes. Tilt angle and rotation were found optimizing to keep the footprint on the primary within a constant $150\, \rm mm$ from the edge of the mirror.}
    \label{fig:imgqual_fnums_prism}
\end{figure*}

Strehl ratios for all 85 cameras are evaluated using the nominal design found in the previous section. Figure \ref{fig:strehls_fnumbs} (top) shows Strehl ratios for cameras 1, 31 and 83. Figure \ref{fig:imgqual_fnums_prism} (top left) shows that 81 cameras are completely diffraction-limited (Strehl ratio greater than 0.8 across the full array)  at $\lambda=1.1\, \rm mm $. At $\lambda = 2 \rm \, mm$, all 85 cameras reach the diffraction limit.

We evaluate f-numbers at the focal plane for the full system consisting of the 85 cameras with their elliptical stops. The f-number  is computed in the time-reverse sense by tracing a cone of marginal rays from the detector focal plane onto L3, to the  stop and finally to a screen that is placed at a large distance from the stop. The solid angle covered by the marginal ray cone is computed fitting an ellipse to the curve drawn by the marginal rays at this far screen. The f-number is computed as $ f/\# = \frac{1}{2 \tan \theta}$, where $\theta$ is obtained as the geometric mean of the semiaxes of this best-fit angular ellipse (the geometric mean preserves the solid angle of the cone as $\Omega = \pi (\sqrt{ab})^2 = \pi \theta^2$). Figure \ref{fig:strehls_fnumbs} (bottom) shows f-numbers for cameras 1, 31 and 83. The f-numbers evaluated over the 85 cameras are shown in Figure \ref{fig:imgqual_fnums_prism} (top right). Figure \ref{fig:imgqual_fnums_prism} (bottom left) shows histograms of f-numbers for a circular stop, demonstrating that elliptical stops achieve a lower f-number than circular stops. Optomechanical design work is under way to determine filter placement from a cryogenical perspective, we leave its impact on f-number as future work.

\begin{figure}[htb]
    \centering
    \includegraphics[trim= 0.01in 0.05in 0in 0.15in, clip, width=\columnwidth]{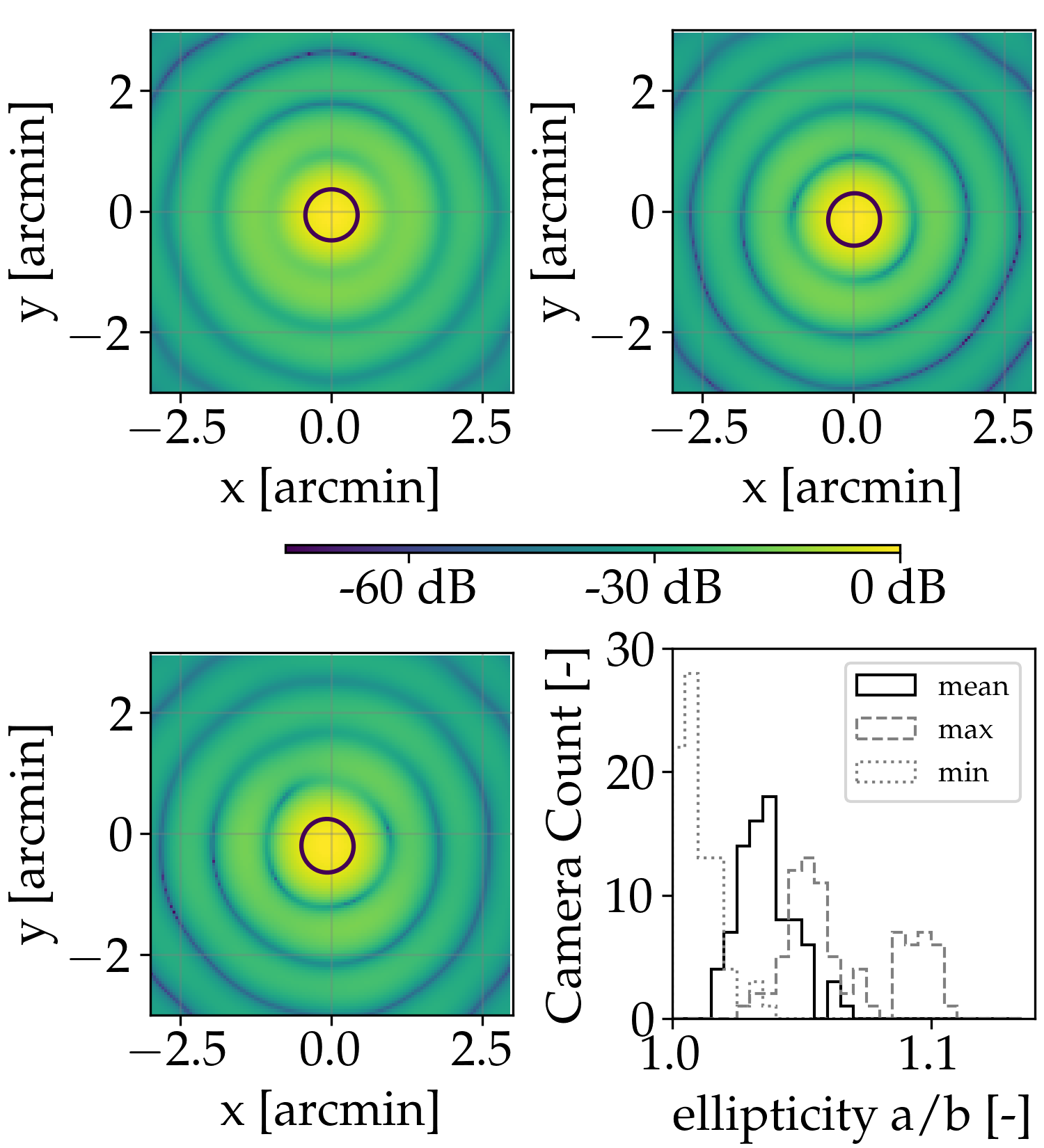}
    \caption{Upper left, upper right and bottom left: Simulated Huygens diffraction beams for 
    camera 1, 31 and 83 respectively for an extreme field located $61 \, \rm mm$
    from the center of the focal plane in the vertical direction. Bottom right: histograms showing the mean, minimum and maximum ellipticity for all 85 cameras.}
    \label{fig:cam_beam_fig}
\end{figure}

Figure \ref{fig:cam_beam_fig} shows beams and ellipticities for three cameras: 1, 31 and 83, with histograms for ellipticities obtained from a Huygens diffraction point spread function model. We evaluate the field of view of each camera by tracing the chief ray from the focal plane to the sky for two extreme focal plane positions in the $y$ direction ($-61 \, \rm{mm}$ and $+61\, \rm{mm}$) corresponding to opposite tips of the hexagon. We compute the dot product of the direction vectors for these two rays. We obtain a per-camera median full field of view of 0.68 degrees, a maximum of 0.71 and a minimum of 0.64 degrees across all cameras.

Telecentricity is evaluated by tracing the chief ray through the system and measuring the angle from the normal for all points in the focal plane. With this procedure we confirm all telecentric angles are lower than 2.5 degrees which is imposed as a hard limit during optimization.

\begin{figure}[h]
    \centering
    \includegraphics[width=0.98\columnwidth]{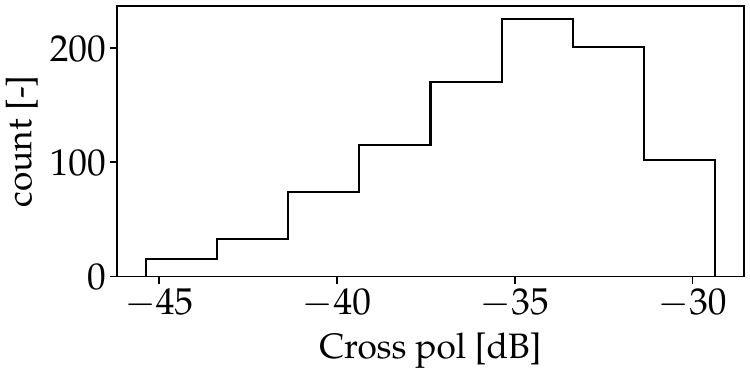}
    \caption{Cross-polarization across the focal plane for the 85 cameras.}
    \label{fig:crosspol}
\end{figure}

Cross polarization is evaluated for the camera and telescope by inserting orthogonal polarizing grids at the detectors and in the far field of the telescope. This is evaluated in 11 fields for the 85 cameras, results are shown in the histogram in Figure \ref{fig:crosspol}. We obtain cross-polarization lower than -29.4 dB in the ray tracing limit.

\begin{figure}[h]
    \centering
    \includegraphics[width=0.99\columnwidth, trim= 0.10in 0.0in 0.0in 0.0in, clip]{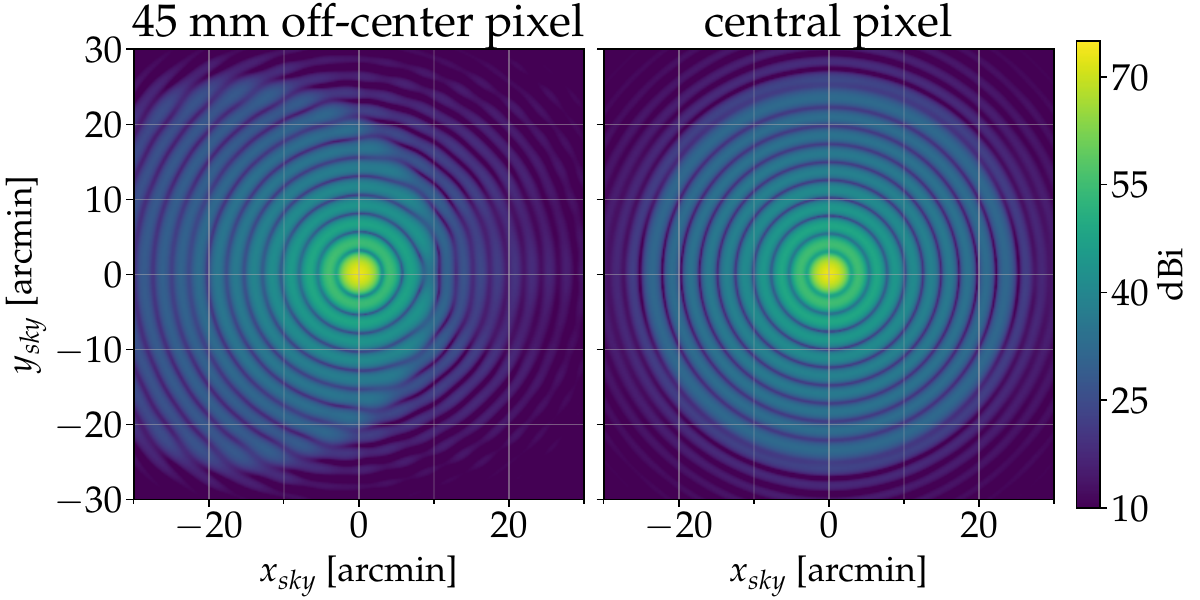}
    \caption{Monochromatic far-field beam shape at 90 GHz obtained with \textsc{Ticra-Tools} (formerly \textsc{GRASP}) by a combination of Physical Optics (for mirrors and stop) and full-wave Method of Moments (for lenses and prism) in time-reverse direction. Illumination is set with a Gaussian beam taper of $-3\, \rm dB$  at 18 degrees. Left panel shows one pixel $45\, \rm mm$ from the center of the detector focal plane, and the right panel shows the central pixel.}
    \label{fig:beamat90}
\end{figure}

The Point Spread Function (PSF) is calculated using an electromagnetic simulation software provided by \textsc{Ticra-Tools} 
(formerly \textsc{GRASP}), in which  surface currents on a reflector are computed when the reflector is both being illuminated by a field and radiating in the surrounding space. 
We use a combination of vector Physical Optics calculations to compute the fields produced by  the mirrors and Method of Moments to compute fields produced by the camera. Physical Optics is a sequential simulation that allows computation of incident and reflected fields on a reflector using an appropriate discretization of the reflector surface and its boundary conditions. 
Method of Moments (MoM) is a full-wave solution of Maxwell equations, which includes internal reflection effects and allows for the application of anti-reflective coatings on refractive surfaces. MoM is used to simulate the three lenses and the prism within one camera. In this simulation the field is propagated in the time-reverse direction starting from 
one pixel of the Camera 1 focal plane, we simulate the central pixel and an off-axis pixel located $45 \, \rm mm$ off-center. The beam pattern of one horn at the detector focal plane is simulated with a Gaussian profile, tapered 3 dB down the peak at 18.1 deg from its boresight, as indicated by an early horn beam model.
An additional stop is placed at the exit of lens 1 (L1) to model radiation absorbed and scattered by the interior of the cryostat (not included in this  sequential analysis). The field at the window towards the tertiary mirror is normalized to have a total irradiated power of $4\pi \, \rm {Watts}$ which gives the beam gain in dBi units. Nominal power spilling past the primary mirror is calculated under $0.5 \rm \%$ for a prism reflectance of 20 dB for the central pixel. A cross polarization lower than 45 dB is obtained relative to  the copolar component. The far field beam shape at $90 \, \rm GHz$ for the $45 \, \rm mm$ off-axis pixel (left) and center pixel (right) are shown in Figure \ref{fig:beamat90}.

\section{Telescope-camera tolerances}
\label{sec:tel_cam_tol}

\begin{figure*}[b]
    \centering
    \begin{subfigure}[t]{0.49\textwidth}
        \centering
        \includegraphics[trim= 0.35in 0.27in 0.35in 0.27in, clip, width=\textwidth]{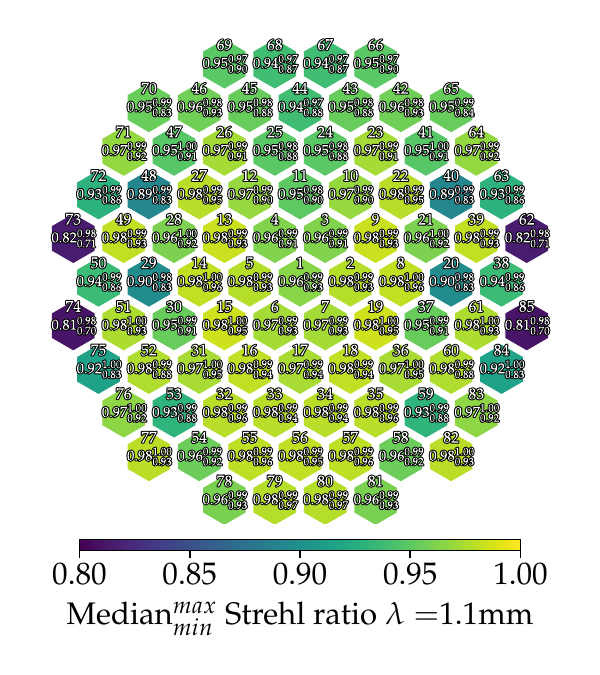}
    \end{subfigure}
    \begin{subfigure}[t]{0.49\textwidth}
        \centering
        \includegraphics[trim= 0.35in 0.27in 0.35in 0.27in, clip, width=\textwidth]{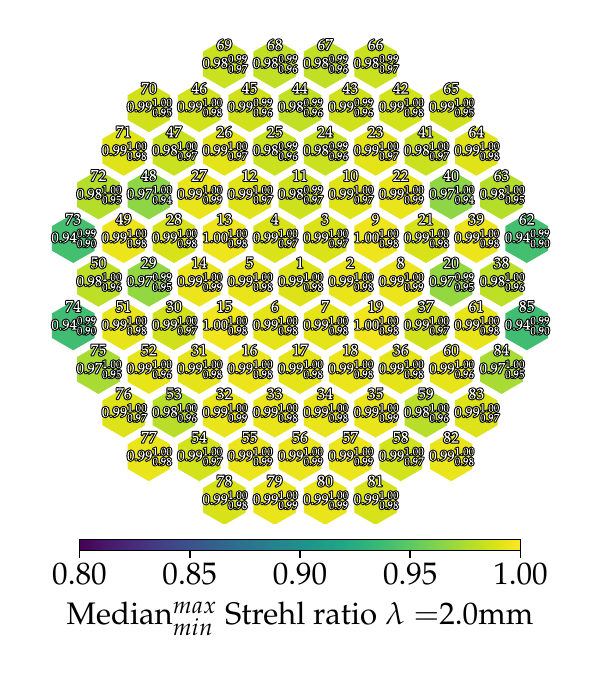}
    \end{subfigure}
    \caption{Left: Strehl ratios at $\lambda=1.1\rm \, mm$ for the 85-camera design presented in section \ref{sec:camera_design} with mirror gravitational and thermal deformations described in section \ref{sec:tolplacements}. Right: The same system evaluated at $2\rm \, mm$ shows all cameras show diffraction-limited performance (above 0.8).}
    \label{fig:deformed_mirror_strehls}
\end{figure*}

\subsection{Camera}

Camera tolerances are obtained with the telescope at its nominal configuration and perturbing the distances between lenses and Lyot stop (4 variables), the lens centering (3 variables) and lens tilts (6 variables). We define a merit function in camera number 8, which computes the average Sterhl ratio in 25 fields. We perturb each variable while keeping the rest at their nominal value, and we search for a $5\%$ degradation in average Strehl. Figure \ref{fig:cam_tolerancing} (two top rows) shows the average Strehl metric as a function of a single variable perturbation in the nominal design. Single variable tolerances are in the $5\, \rm mm$ range and allowable tilts are in the $4 \, \rm {degree}$ range using this criterion. Multivariable Monte Carlo analysis returns a joint allowable tolerance of $1.5\, \rm mm$ mm in lens distances and $3.5\, \rm mm$ in lens decenterings, with a $1.5 \, \rm degree$ tolerance in lens tilt. Table \ref{tab:camera_telescope_tol} shows a summary of this tolerancing analysis and Figure \ref{fig:cam_tolerancing} (bottom panel) shows a histogram of average Strehl ratios for independent and uncorrelated Monte Carlo realizations using these values.

\begin{table}
    \centering
    \begin{tabular}[h]{ccc}
        \toprule
         Tolerance type & single par. & joint MC \\
         \midrule{}
        distance [mm]   & 5 & 1.5 \\
        decenter [mm]   & 6 & 3.5 \\
        tilt     [deg]  & 4 & 1.5 \\
        \bottomrule
    \end{tabular}
    \caption{Tolerancing for camera 8. Single parameter values (column single par.) obtained for a degradation of 0.05 in average Strehl ratio across the 120 mm focal plane. Joint Monte Carlo results in the same degradation varying all variables independently.}
    \label{tab:camera_telescope_tol}
\end{table}

\subsection{Deformed mirrors}

We evaluate Strehl ratios with the three deformed mirrors. We include thermal warping and gravitational deformations as described in Section \ref{sec:tel_nominalperformance} \ref{sec:tolplacements}, we also allow the camera to vary position to refocus. We obtain acceptable image quality as characterized by Strehl ratio coverage. In particular, we find that 81 cameras have diffraction-limited image quality (Strehl ratio > 0.8) at $\lambda=1.1\,\rm mm$ as shown in Figure \ref{fig:deformed_mirror_strehls} (left) and at $\lambda = 2\,\rm mm$ we find that diffraction-limited performance is achieved in all 85 cameras as shown in Figure \ref{fig:deformed_mirror_strehls} (right).

%%%%%%%%%%%%%%%%%%%%%%%%%%  end body  %%%%%%%%%%%%%%%%%%%%%%%%%%

%%%%%%%%%%%%%%%%%%%%%%%%% conclusion %%%%%%%%%%%%%%%%%%%%%%%
\section{Conclusion}
\label{sec:conclusion}

We have presented a freeform three-mirror anastigmatic large-aperture telescope, capable of observing a wide 9.4 degree field of view at $\lambda = 1.1\, \rm mm$. A detailed description of this design is given accompanied by a suite of metrics that indicate excellent performance. We have also presented a detailed description of an array of 85 cameras that can observe up to $\lambda=1.1\, \rm mm$ in wavelength with diffraction-limited performance in 81 of 85 cameras and with all 85 cameras at $2\, \rm mm$. We have shown expected performance and tolerancing. We continue to iterate on the opto-mechanical design in order to prepare this design for manufacture.

\section{Appendix}

\subsection{Telescope and camera tolerancing}
\begin{figure}[t]
    \centering
    \includegraphics[width=\columnwidth, trim={0 0 0 0}, clip]{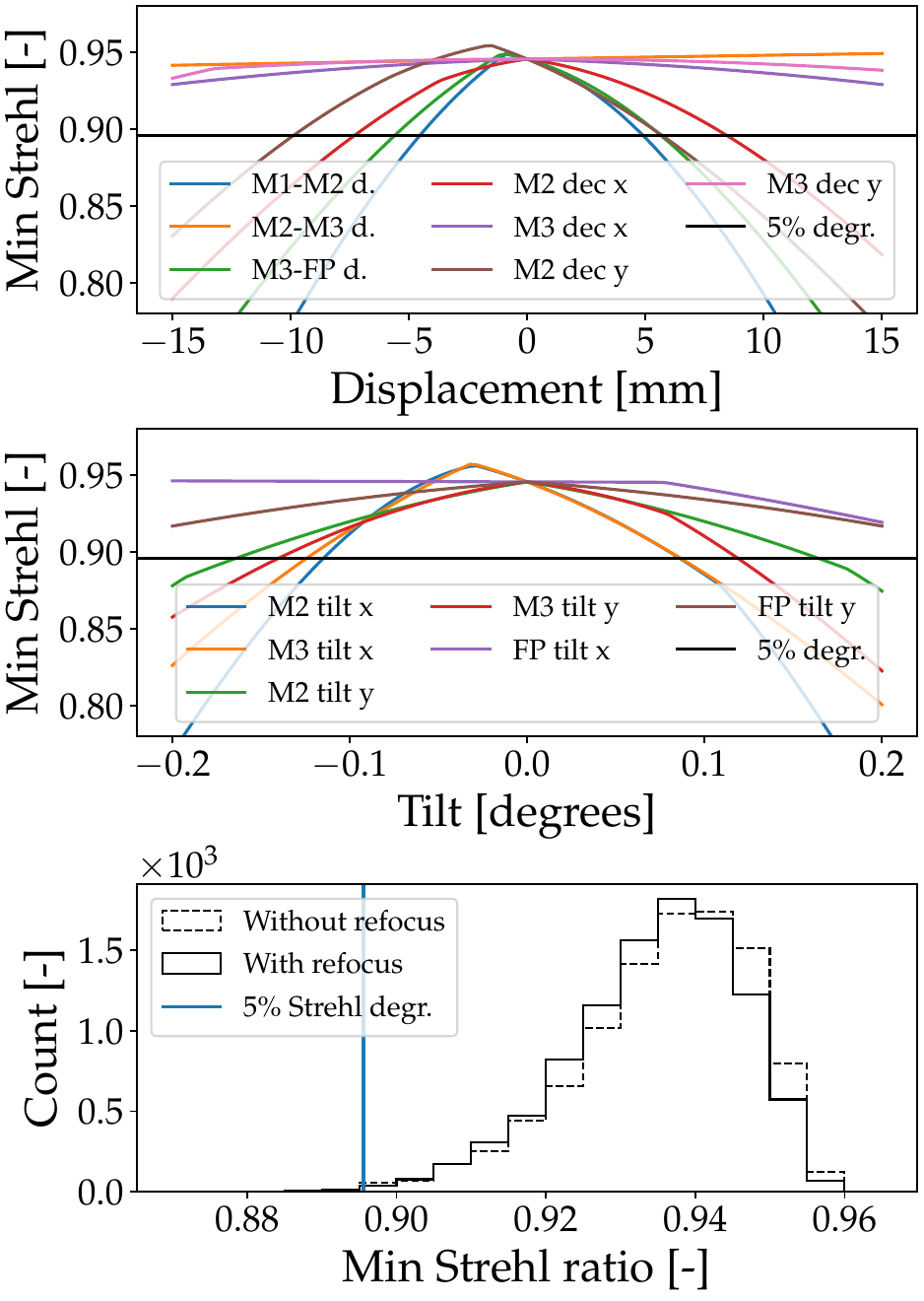} % see sensitivity_sweep folder
    \caption{Tolerancing of the telescope three-mirror system without reimaging camera optics.
    Top: Single parameter variations in the minimum Strehl ratio evaluated at the edge of the focal plane 4.6 degrees from the boresight.
    Center: Single parameter variations for the tilt parameters of the minimum Strehl ratio.
    Bottom: Monte Carlo tolerancing simulation of the three-mirror telescope presented in this work.  Histogram describes the Strehl ratio variation, during random changes to mirror misplacement and tilts as described in Section \ref{sec:tolplacements} and for parameters shown in Table \ref{tab:tel_tols}.}
    \label{fig:tol}
\end{figure}

\begin{figure}[t]
    \centering
	\includegraphics[width=\columnwidth]{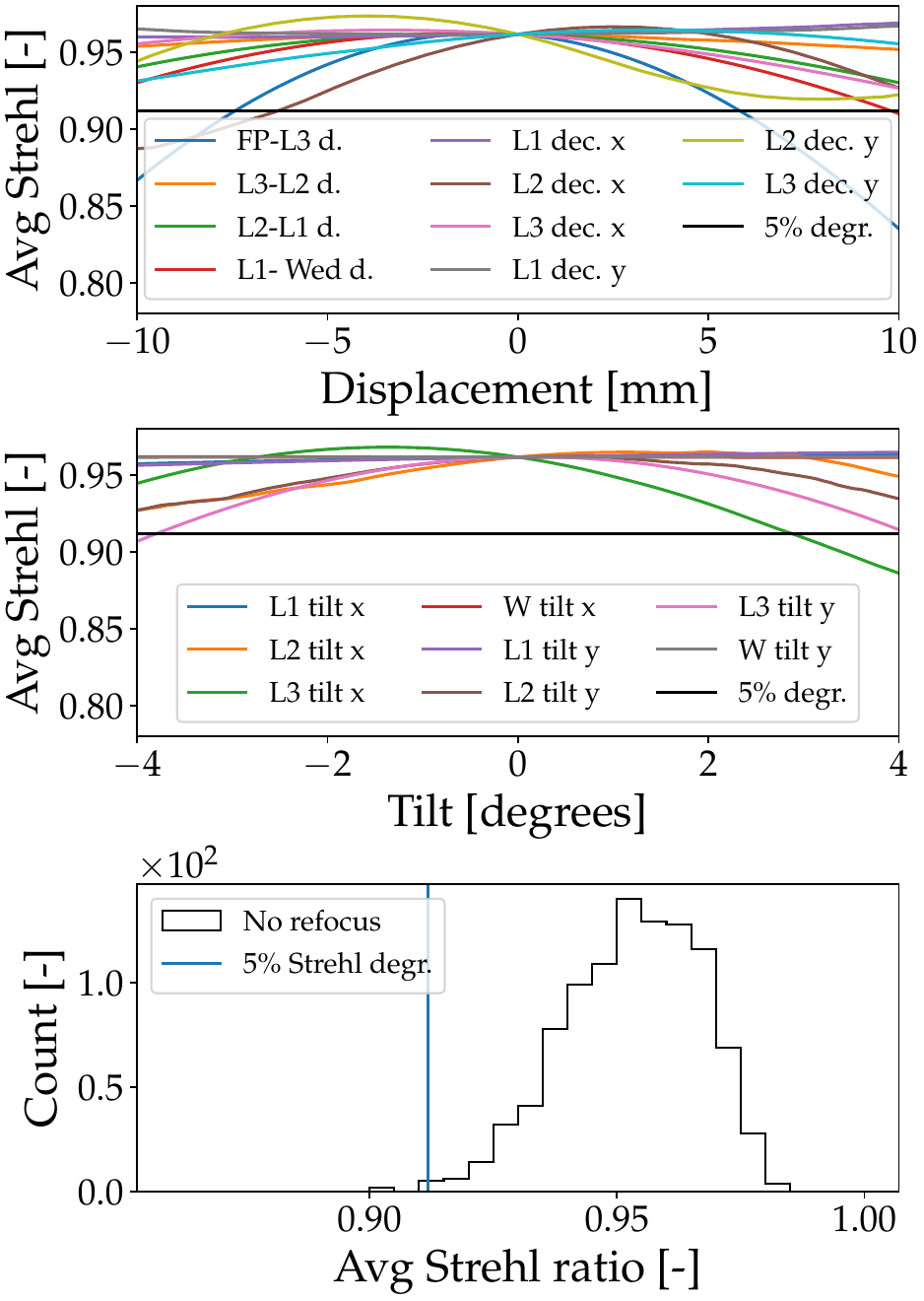}
    \caption{Single variable tolerancing for camera 8. Top: Average Strehl ratio for a perturbation in lens distances and decenter. Center: Average Strehl for a perturbation in lens tilts. Bottom: Monte Carlo sampled average Strehl ratios for the values in Table \ref{tab:camera_telescope_tol}.}
    \label{fig:cam_tolerancing}
\end{figure}

In Section \ref{sec:tel_tol} and Section \ref{sec:tel_cam_tol} we describe the merit function used to compute tolerancing metrics. Figure \ref{fig:tol} and Figure \ref{fig:cam_tolerancing} show the merit function variation under single camera variations for displacement variables (top panel), angular tilt variables (center panel) and a histogram of Monte Carlo realizations for the case of the telescope alone (Figure \ref{fig:tol}) and  the three-mirror telescope system with camera 8 (Figure \ref{fig:cam_tolerancing}).

\begin{backmatter}
\bmsection{Funding}
Placeholder for funding (will be replaced by journal).
\bmsection{Acknowledgments}
We acknowledge the valuable contribution of Richard Hills (1945-2022 \cite{longair2023richard}) to the optical design and optimization of the freeform three-mirror anastigmat presented here. We will sorely miss him. We acknowledge FONDEF ID21I10236 for supporting the licensing of the TICRA software, and the Geryon cluster at the Centro de Astro-Ingenieria UC, which was used for the full-wave calculations in this paper. BASAL CATA PFB-06, the Anillo ACT-86, 
FONDEQUIP AIC-57, and QUIMAL 130008 provided funding the Geryon cluster. We thank Stig Busk S{\o}rensen and the whole TICRA staff for their valuable help building the Grasp model. This document was prepared by the CMB-S4 collaboration using the resources of the Fermi National Accelerator Laboratory (Fermilab), a U.S. Department of Energy, Office of Science, HEP User Facility. Fermilab is managed by Fermi Research Alliance, LLC (FRA), acting under Contract No. DE-AC02-07CH11359. JEG acknowledges support from the Swedish National Space Agency (SNSA/Rymdstyrelsen) and the Swedish Research Council (Reg. no. 2019-03959). JEG also acknowledges support from the European Union (ERC, CMBeam, 101040169).
\bmsection{Disclosures} The authors declare no conflicts of interest.
\bmsection{Data availability} The optical design presented here can be reproduced using the tables in this document.
\end{backmatter}

%%%%%%%%%%%%%%%%%%%%%%% References %%%%%%%%%%%%%%%%%%%%%%%%%

\bibliography{biblio}
\end{document}